# Time-resolved dynamic CBCT reconstruction using prior-model-free spatiotemporal Gaussian representation (PMF-STGR)

## Running Title:
## Dynamic CBCT Reconstruction Via Spatiotemporal Gaussians

Jiacheng Xie
Hua-Chieh Shao
You Zhang

*The Advanced Imaging and Informatics for Radiation Therapy (AIRT) Laboratory*
*The Medical Artificial Intelligence and Automation (MAIA) Laboratory*
*Department of Radiation Oncology, University of Texas Southwestern Medical Center, Dallas, TX 75390, USA*

Corresponding address:

You Zhang
Department of Radiation Oncology
University of Texas Southwestern Medical Center
2280 Inwood Road
Dallas, TX 75390
Email: You.Zhang@UTSouthwestern.edu
Tel: (214) 645-2699

**Abstract**

*Objective.* Time-resolved CBCT imaging, which reconstructs a dynamic sequence of CBCTs reflecting intra-scan motion (one CBCT per x-ray projection without phase sorting/binning), is highly desired for regular/irregular motion characterization, patient setup, and motion-adapted radiotherapy. Representing patient anatomy and associated motion fields as 3D Gaussians, we developed a Gaussian representation-based framework (PMF-STGR) for fast and accurate dynamic CBCT reconstruction. *Approach.* PMF-STGR comprises three major components: a dense set of 3D Gaussians to reconstruct a reference-frame CBCT for the dynamic sequence; another 3D Gaussian set to capture three-level, coarse-to-fine motion-basis-components (MBCs) to model the intra-scan motion; and a CNN-based motion encoder to solve projection-specific temporal coefficients for the MBCs. Scaled by the temporal coefficients, the learned MBCs will combine into deformation vector fields (DVFs) to deform the reference CBCT into projection-specific, time-resolved CBCTs to capture the dynamic motion. Due to the strong representation power of 3D Gaussians, PMF-STGR can reconstruct dynamic CBCTs in a 'one-shot' training fashion from a standard 3D CBCT scan, without using any prior anatomical/motion model. *Main results.* We evaluated PMF-STGR using XCAT phantom simulations and real patient full/half-fan scans. Metrics, including the image relative error (RE), structural-similarity-index-measure (SSIM), tumor center-of-mass-error (COME), and landmark localization error (LE), were used to evaluate the accuracy of solved dynamic CBCTs and motion. PMF-STGR shows clear advantages over a state-of-the-art, implicit neural representation (INR)-based approach, PMF-STINR. Compared with PMF-STINR, PMF-STGR reduces reconstruction time by ~50% while reconstructing less blurred images with comparable/better motion accuracy. For XCAT, the mean(±s.d.) RE, SSIM, and COME were 0.128(0.009), 0.990(0.002), and 0.71mm(0.40mm) for PMF-STGR, compared with 0.149(0.016), 0.944(0.006), and 0.94mm(0.18mm) for PMF-STINR. For patients, the mean(±s.d.) landmark LE was 1.40mm(0.34mm) for PMF-STGR, and 1.54mm(0.35mm) for PMF-STINR. *Significance.* With improved efficiency/accuracy, PMF-STGR enhances the applicability of dynamic CBCT imaging for potential clinical translation.

## 1. Introduction

In radiotherapy, cone-beam computed tomography (CBCT) is widely used in clinical practice, providing volumetric imaging with excellent spatial resolution as guidance for patient setup, treatment verification, and plan adaptation (Jaffray *et al.*, 2002; Oldham *et al.*, 2005). Due to the prolonged acquisition time, patient motion, primarily respiratory motion (with a cycle of 3–5 seconds), traditional 3D CBCT imaging introduces artifacts and blurring in the reconstructed images (Rit *et al.*, 2011). To mitigate artifacts and capture motion more accurately, four-dimensional (4D) CBCT was developed as the current clinical standard (Abulimiti *et al.*, 2023; Sonke *et al.*, 2005). 4D-CBCT sorts projections into predefined motion bins and reconstructs semi-static CBCT images for each bin to approximate an average motion pattern, with an underlying assumption that anatomical motion is periodic and regular, which is often inaccurate (Yasue et al., 2021). Although patient-specific prior-image-guided approaches, ranging from iterative prior-regularized optimization (PRIOR) (Hu *et al.*, 2022) to the recent deep-learning framework DPI-MoCo (Hu *et al.*, 2025), have greatly reduced streaking and respiratory-motion artifacts in 4D-CBCT, the technique's temporal resolution is still fundamentally constrained by phase-sorting/binning, which compresses the breathing curve into only a few (<=10) discrete motion states. As a result, 4D-CBCT fails to capture time-resolved irregular motion, potentially affecting patient setup and dose delivery accuracy (Clements et al., 2013; Li et al., 2018). Additionally, motion sorting typically relies on surrogates (e.g., surface markers), which can introduce errors due to the limited correlation between surrogate motion and internal anatomy (Yan et al., 2008). A key approach to overcoming the limitations of 4D-CBCT is reconstructing a time-resolved dynamic sequence of CBCTs, which yields a CBCT for each x-ray projection, offering the ultimate spatial and temporal resolutions to capture intra-scan dynamic motion. In radiotherapy, dynamic CBCTs are ideal for visualizing moving patient anatomy for treatment planning and optimizing motion management strategies during pre-treatment, reconstructing dynamic doses and determining the real accumulated dose during treatment (Zou et al., 2014), and guiding plan adaptation of future treatments (Brock, 2019). Despite these advantages, time-resolved dynamic CBCT is not yet clinically available due to its reconstruction challenge. Conventional CBCT reconstruction requires hundreds of projections (Feldkamp et al., 1984), whereas a single 2D projection lacks sufficient information for accurate dynamic CBCT reconstruction.

Several studies have attempted dynamic CBCT reconstruction via modeling dynamic anatomy/motion in a simplified manner, such as using low-rank factorization (Cai *et al.*, 2014) or as linear combinations of basis images (Gao *et al.*, 2018), but these approaches may struggle to capture complex 3D motion and generalize beyond regular breathing patterns. Unlike these reconstruction-based methods, deformation-driven approaches attempted to reconstruct dynamic CBCTs by integrating prior knowledge of the patient anatomy and/or the motion. To satisfy the extreme under-sampling scenario of single-projection based dynamic CBCT reconstruction, motion models based on principal component analysis (PCA)-assisted dimensionality reduction were developed using patient-specific prior 4D-CTs (Li *et al.*, 2011a; Wei *et al.*, 2020), on top of an anatomical model extracted as one phase image of the prior 4D-CT set. A key limitation of prior-model-based dynamic CBCT reconstruction is the assumption of an invariant anatomical model, which may not hold due to non-deformation-related changes such as contrast enhancement, tissue inflammation, or disease progression (Zhang *et al.*, 2017). Additionally, using an anatomical model from a different imaging system (simulation CT scanner) rather than the same CBCT device can introduce discrepancies due to variations in energy, scatter, noise, and image intensity (Zhang *et al.*, 2015). The assumption of a stable motion model also fails to account for inter-fractional deformation and motion pattern changes (Zhang *et al.*, 2013). Moreover, generating such a model from prior 4D-CTs can be challenging, as motion sorting artifacts may be present, and not all patients have prior 4D-CT data available. Existing methods are often limited to simulation studies using simplified geometries or scan conditions, which may not generalize to real-world CBCT acquisition conditions and patient-specific motion variations. Besides the pure reconstruction-driven or deformation-driven approaches, Huang et al. introduced a surrogate-driven respiratory motion model (SuPReMo) to reconstruct dynamic CBCTs from unsorted projection data via a motion-compensated strategy (Huang *et al.*, 2024). This method relies on motion surrogate signals, combined with B-splines, to capture intra-treatment motion fields. Specifically, two surrogate signals are obtained from projection images and filtered to remove background intensity variations before feeding into the motion model. Although showing promising results on simulated and real patient data, the model faces limitations: a consistent surrogate may not be extractable from all projections due to limited fields of view. Consequently, the accuracy of SuPReMo heavily depends on the quality of these surrogate signals. A recent PMF-STINR (Shao *et al.*, 2024) study introduced a 'one-shot' solution for dynamic CBCT reconstruction using conventional 3D CBCT scans without relying on prior modeling, motion sorting/binning, or surrogate signals, leveraging the capability of implicit neural representation (INR) learning implicit mappings of complex 3D scenes from sparse 2D views (Mildenhall *et al.*, 2021). PMF-STINR reconstructs dynamic CBCT by combining three components: a spatial INR, a temporal INR, and a learnable cubic B-spline motion model. The spatial INR reconstructs a reference-frame CBCT, while the temporal INR, together with the B-spline-based motion model, estimate time-dependent motion fields relative to the reference-frame CBCT. The motion model learns basis motion patterns directly from to-be-reconstructed projections, and the temporal INR captures their time-varying weights. By parameterizing basis motion patterns on a coarser control-point grid using B-splines, PMF-STINR effectively models smooth, dense motion fields in a data-driven manner. Evaluated on both simulated digital phantoms and real patient cone-beam projections, this approach demonstrated state-of-the-art performance, offering a significant advancement over traditional 4D-CBCTs. However, a notable limitation of PMF-STINR is its long training time, often exceeding 3 hours on V100 (~80 minutes on RTX 4090) to reconstruct a dynamic full-fan CBCT sequence. Its memory consumption is also considerable, necessitating the use of large-memory GPUs and preventing the direct reconstruction of high-resolution CBCTs ($1 \times 1 \times 1$ mm$^3$, for instance) (Shao *et al.*, 2024). Also, due to the limitation of INR in representing high-frequency signals, the images reconstructed by PMF-STINR appear blurred in sharp-transition regions (e.g. bony areas) and need special multi-resolution reconstructions for partial mitigation. In addition, using a learnable B-spline-based interpolant to represent the motion model, PMF-STINR allows smooth motion representation but introduces errors in areas with discontinuous sliding motion (Loring *et al.*, 2005; Al-Mayah *et al.*, 2009).

Recently, a machine learning technique named 3D Gaussian splatting (3DGS) (Kerbl *et al.*, 2023) has emerged for computer vision applications including view synthesis and image reconstruction. It represents scenes using a collection of 3D Gaussian functions instead of dense voxel grids or neural networks, preserving the continuous volumetric properties of studied objects/fields while avoiding unnecessary computations in empty space, which allows for significantly faster rendering speeds compared to INR-based methods (Mildenhall *et al.*, 2021). Building on static 3D scene reconstruction, several efforts have been made to extend Gaussian splatting to dynamic scenes. Wu et al. incorporated a deformation network into 3D Gaussian splatting, enabling real-time dynamic view synthesis (Wu *et al.*, 2023). Luiten et al. modeled dynamic scenes by allowing Gaussians to move and rotate over time while maintaining consistent properties such as color, opacity, and size, achieving both dynamic novel-view synthesis and motion tracking (Luiten *et al.*, 2024). However, the above works are all developed for natural light imaging. To accommodate for X-ray imaging, Zha et al.

proposed R2-Gaussian (Zha *et al.*, 2024) for tomographic sparse-view reconstruction with tailored Gaussian rendering and voxelization for X-ray imaging. By incorporating a deformation network into R2-Gaussian, a recent work from Fu et al. (Fu *et al.*, 2025) proposed an end-to-end framework for 4D-CBCT reconstruction using 4D Gaussian representation. While achieving good reconstruction accuracy, the method relies on phase sorting and binning, preventing time-resolved dynamic CBCT reconstruction to resolve irregular breathing patterns. It lacks explicit modeling of true anatomical motion through conventional deformation vector fields (DVFs). Instead, the motion is represented through deforming 3D Gaussian parameters—position, scale, rotation, and density—via a multi-head multi-layer perceptron (MLP) decoder guided by the encoded spatiotemporal features. Compounding intensity changes with motion (due to Gaussian kernel summation), the motion fields of the Gaussians do not represent true physical motion. This representation, though effective for image synthesis, diverges from the DVF-based motion models essential for motion management and guidance in radiotherapy, where accurate, interpretable, and physics-based motion fields are critical for dose accumulation (Yan *et al.*, 1999; Keall *et al.*, 2005; Rietzel *et al.*, 2005) and contour propagation (Rietzel and Chen, 2006; Wang *et al.*, 2008; Xie *et al.*, 2008). Additionally, its long training time (~3 hours) hinders clinical practicality, especially for adaptive radiotherapy.

To address the remaining issues of PMF-STINR, we propose in this study a prior-model-free spatiotemporal Gaussian representation (PMF-STGR) approach, which uses the strong representation power of 3D Gaussians to reconstruct dynamic CBCTs. PMF-STGR consists of three key components: a dense 3D Gaussian set to reconstruct a reference-frame CBCT with fine details, another Gaussian set to model intra-scan motion through three-level, coarse-to-fine motion-basis components (MBCs) to capture voxelwise motion pattern variations, and a CNN-based motion encoder that computes projection-specific temporal coefficients for these MBCs. The coefficient-scaled MBCs are combined into DVFs to deform the reference CBCT into projection-specific CBCTs, capturing dynamic motion. Similar to PMF-STINR, PMF-STGR enables an 'one-shot' dynamic CBCT reconstruction from a standard 3D CBCT scan, eliminating the need for prior anatomical or motion models. Furthermore, leveraging the strong representation power of 3D Gaussians, PMT-STGR reconstructs more accurate dynamic CBCTs, which provide better image details and more accurate motion characterization, while with reduced computation time and memory cost. We evaluated PMF-STGR using XCAT phantom simulations and real patient scans, with XCAT simulating lung CBCTs under seven free-breathing scenarios with varying motion irregularities. For real patients, CBCT projection sets from eight patient cases were used. Reconstruction accuracy was assessed using relative error (RE), structural similarity index measure (SSIM), tumor center-of-mass error (COME), and landmark localization error (LE). Compared with PMF-STINR, PMF-STGR reduces reconstruction time by ~50% (~40 mins on RTX 4090 for a full-fan scan), while reconstructing less blurred images with comparable/better motion accuracy. With improved efficiency/accuracy, PMF-STGR enhances the applicability of dynamic CBCT imaging for potential clinical translation.

## 2. Materials and Methods

*2.1 Dynamic CBCT reconstruction overview*

The dynamic CBCT reconstruction is typically formulated as an optimization problem:

$$\{\hat{\boldsymbol{I}}(\boldsymbol{x},p)\} = argmin_{\{\boldsymbol{I}(\boldsymbol{x},p)\}}(|\mathcal{P}\{\boldsymbol{I}(\boldsymbol{x},p)\} - \{p\}|^2 + \lambda\mathcal{R}), \quad (1)$$

where $\boldsymbol{P}$ denotes a consecutive sequence of cone-beam x-ray projections, and $p \in \boldsymbol{P}$ denotes one of the projections from the projection set. $\boldsymbol{I}(\boldsymbol{x},p)$ represents the linear attenuation coefficients (isotropic density) at spatial coordinates $\boldsymbol{x} \in \mathbb{R}^3$, or equivalently the to-be-solved dynamic CBCT volume, corresponding to the projection $p$. $\mathcal{P}$ denotes the projection matrix, and $\lambda$ is the weighting factor of the regularization term $\mathcal{R}$. Solving the highly ill-posed optimization problem in Eq.1 can be extremely challenging as the dynamic sequence $\boldsymbol{I}(\boldsymbol{x},p)$ can contain $\mathcal{O}(10^8)$ or more voxels to reconstruct, given the 2D projection set $\boldsymbol{P}$. To simplify the inverse problem, we fit Eq. 1 into a motion-compensated reconstruction framework, by solving a reference-frame CBCT $\boldsymbol{I}_{\text{ref}}(\boldsymbol{x})$ and the intra-scan motion with respect to $\boldsymbol{I}_{\text{ref}}(\boldsymbol{x})$. The de-coupling of anatomy ($\boldsymbol{I}_{\text{ref}}(\boldsymbol{x})$ ) and motion assumes that excluding physiological motion, the underlying anatomy remains unchanged during the scan, which is generally true considering the time scale of a CBCT scan (~ 1 min). By deforming the reference CBCT with a sequence of time-dependent DVFs $\boldsymbol{d}(\boldsymbol{x},p)$, the dynamic CBCT sequence $\boldsymbol{I}(\boldsymbol{x},p)$ can be obtained as:

$$\boldsymbol{I}(\boldsymbol{x},p) = \boldsymbol{I}_{\text{ref}}\big(\boldsymbol{x} + \boldsymbol{d}(\boldsymbol{x},p)\big). \quad (2)$$

For further dimension reduction to address the ill-posed problem, each time-dependent motion field $\boldsymbol{d}(\boldsymbol{x},p)$ can be approximated (Zhao *et al.*, 2012) by a summation of products of spatial ($\boldsymbol{e}_i(\boldsymbol{x})$) and temporal ($\boldsymbol{w}_i(p)$) components:

$$\boldsymbol{d}(\boldsymbol{x},p) = \sum_{i=1}^{3} \boldsymbol{w}_i(p) \times \boldsymbol{e}_i(\boldsymbol{x}). \quad (3)$$

The spatial component $\boldsymbol{e}_i(\boldsymbol{x})$ serves as a set of basis functions that span the motion space, capturing various motion patterns. The temporal component $\boldsymbol{w}_i(p)$ represents the coefficients obtained from time-varying projections that map the contribution of each spatial basis component to describe the intra-scan dynamic motion. By decoupling the $\boldsymbol{d}(\boldsymbol{x},p)$ as a linear combination of spatial component weighted by their corresponding temporal coefficients, we achieve a low-rank approximation that further reduces the unknowns in the ill-posed problem. Thus, dynamic CBCT imaging is equivalent to reconstructing a reference CBCT $\boldsymbol{I}_{\text{ref}}$, while determining time-varying linear weightings $\boldsymbol{w}_i(p)$ of motion basis components (MBCs) $\boldsymbol{e}_i(\boldsymbol{x})$ that capture the underlying anatomical motion. Following the previous study (Shao *et al.*, 2024), we used three MBCs (i.e. $i = 1, 2, 3$) for each Cartesian direction to describe complex breathing motion. In prior studies (Li *et al.*, 2011b; Zhang *et al.*, 2013; Wei *et al.*, 2020; Zhang *et al.*, 2023), the reference-frame volume $\boldsymbol{I}_{\text{ref}}$ and/or the motion model $\boldsymbol{e}_i(\boldsymbol{x})$ are usually derived from prior 4D-CT/CBCT scans, introducing uncertainties due to anatomical and motion pattern variations between prior and new imaging sessions. In our prior-model-free framework, we aim to solve $\boldsymbol{I}_{\text{ref}}$, $\boldsymbol{w}_i(p)$, and $\boldsymbol{e}_i(\boldsymbol{x})$ solely from each projection set, yielding a 'one-shot' approach for robust learning.

*2.2 Gaussian representation and splatting*

In the framework of radiative Gaussians (Zha *et al.*, 2024), the target objects (for instance, a to-be-reconstructed CBCT) are modeled with a set of learnable 3D kernels $\mathbb{G}^3 = \{G_i^3\}_{i=1,\dots,M}$ such that each kernel $G_i^3$ defines a local Gaussian-shaped density field:

$$G_i^3(\boldsymbol{x} \mid \rho_i, \boldsymbol{p}_i, \boldsymbol{\Sigma}_i) = \rho_i \cdot \exp\left(-\frac{1}{2}(\boldsymbol{x} - \boldsymbol{p}_i)^\top \boldsymbol{\Sigma}_i^{-1}(\boldsymbol{x} - \boldsymbol{p}_i)\right), \tag{4}$$

where $\rho_i$ , $\boldsymbol{p}_i \in \mathbb{R}^3$ and $\boldsymbol{\Sigma}_i \in \mathbb{R}^{3\times3}$ are learnable parameters representing central density, position, and covariance, respectively. The overall density $\sigma(\boldsymbol{x})$ at $\boldsymbol{x} \in \mathbb{R}^3$ can be obtained by summing the densities of kernels, which is the voxelization operation for Gaussians:

$$\sigma(\boldsymbol{x}) = \sum_{i=1}^{M} G_i^3(\boldsymbol{x} \mid \rho_i, \boldsymbol{p}_i, \boldsymbol{\Sigma}_i). \tag{5}$$

In X-ray imaging, the pixel value $I(\boldsymbol{r})$ along an X-ray $\mathbf{r}(t) = \mathbf{o} + t\mathbf{d} \in \mathbb{R}^3$, the path $(t)$ of which bounded by $t_n$ and $t_f$, is represented by the integral of density:

$$I(\boldsymbol{r}) = \int_{t_n}^{t_f} \sigma(\boldsymbol{r}(t))dt, \tag{6}$$

where $\mathbf{o}$ is the X-ray source, and $\mathbf{d}$ is the unit vector pointing from the source to the detector. In the context of Gaussians, by substituting Eq. 5 with Eq. 6:

$$I(\boldsymbol{r}) = \sum_{i=1}^{M} \int G_i^3(\boldsymbol{r}(t) | \rho_i, \boldsymbol{p}_i, \boldsymbol{\Sigma}_i)dt, \tag{7}$$

where $I(\boldsymbol{r})$ represents the rendered pixel value. This allows the integration of each 3D Gaussian independently to rasterize an X-ray projection. To approximate a cone-beam X-ray scanner in ray space, the local affine transformation is applied to Eq. 7, yielding:

$$I(\boldsymbol{r}) \approx \sum_{i=1}^{M} \int G_i^3\left(\tilde{\boldsymbol{x}} \middle| \rho_i, \underbrace{\phi(\boldsymbol{p})}_{\tilde{\boldsymbol{p}}_i}, \underbrace{\boldsymbol{J}_i \boldsymbol{W} \boldsymbol{\Sigma}_i \boldsymbol{W}^\top \boldsymbol{J}_i^\top}_{\tilde{\boldsymbol{\Sigma}}_i}\right) dx_2, \tag{8}$$

where $\tilde{\boldsymbol{x}} = [x_0, x_1, x_2]^\top$ represents a point in the ray space, $\tilde{\boldsymbol{p}}_i \in \mathbb{R}^3$ is the new Gaussian position by applying the projective mapping $\phi$, and $\tilde{\boldsymbol{\Sigma}}_i \in \mathbb{R}^{3\times3}$ is the new Gaussian covariance with perspective to the local approximation matrix $\boldsymbol{J}_i$ and viewing transformation matrix $\boldsymbol{W}$. The projective mapping $\phi$, local approximation matrix $\boldsymbol{J}_i$, and viewing transformation matrix $\boldsymbol{W}$ are determined by scanner parameters. Through projection, the 3D Gaussian distribution is transformed to a 2D Gaussian distribution:

$$I(\boldsymbol{r}) = \sum_{i=1}^{M} G_i^2\left(\hat{\boldsymbol{x}} \,\middle|\, \underbrace{\sqrt{\frac{2\pi|\tilde{\Sigma}_i|}{|\hat{\Sigma}_i|}}\rho_i}_{\hat{\rho_i}}, \hat{\boldsymbol{p}}_i, \hat{\boldsymbol{\Sigma}}_i\right), \tag{9}$$

where $\hat{\boldsymbol{x}} \in \mathbb{R}^2$, $\hat{\boldsymbol{p}} \in \mathbb{R}^2$, and $\hat{\boldsymbol{\Sigma}} \in \mathbb{R}^{2\times 2}$ are derived by dropping the third rows and columns of $\tilde{\boldsymbol{x}}$, $\tilde{\boldsymbol{p}}_i$, and $\tilde{\boldsymbol{\Sigma}}_i$, respectively. By summing up these 2D Gaussians, an X-ray projection can be quickly generated from 3D Gaussian-represented CBCT volumes via the 'Gaussian splatting' process, allowing iterative reconstructions to be performed.

Based on the foundation of the PMF-STINR framework, by PMFT-STGR, we used 3D Gaussians to represent the reference-frame CBCT $\boldsymbol{I}_{\text{ref}}$ (instead of INRs) to better capture the fine details of the anatomy in a more efficient fashion. In addition, we also used 3D Gaussians to represent the MBCs (instead of B-spline interpolants) for the motion model learning, to better capture the spatial variations of the motion patterns and avoid over-smoothing. Following a recent work (Shao *et al.*, 2025), we also replaced the temporal INRs of PMF-STINR with a CNN-based motion encoder, to learn time-dependent motion coefficients that scale MBCs to yield dynamic DVFs. Such a motion encoder allows the learned anatomy and motion model to be applied towards subsequent real-time motion monitoring.

## *2.3 The PMF-STGR method*

### *2.3.1 Overview of PMF-STGR*

Figure 1 illustrates the workflow of the PMF-STGR framework, which comprises three primary components: reference-frame CBCT Gaussians, a CNN-based motion encoder, and MBC Gaussians. The reference-frame CBCT Gaussians are employed to reconstruct a reference-frame image $\boldsymbol{I}_{\text{ref}}$, representing the patient anatomy. The MBC Gaussians model the data-driven motion basis components $\boldsymbol{e_i}(\boldsymbol{x})$ directly learned from the cone-beam projections. Concurrently, the CNN-based motion encoder determines the time(projection)-varying temporal coefficients $\boldsymbol{w}_i(p)$ corresponding to these MBCs. By integrating the resolved temporal coefficients with the MBC Gaussians, a motion sequence is established that characterizes the time-dependent DVFs of the dynamic CBCT sequence relative to the reference CBCT.

The reference-frame CBCT Gaussians reconstruct the image $\boldsymbol{I}_{\text{ref}}$ by optimizing the parameters of a Gaussian distribution. To derive an X-ray projection $p$ from these Gaussians, one can employ the Gaussian splatting-based X-ray rasterizer as described in Eq. 9. Alternatively, a CUDA-based Gaussian voxelizer (Eq. 5) can be utilized to transform the Gaussian representation into a 3D voxelized volume, which can then be processed using a voxel-based cone-beam projector, such as the Operator Discretization Library (ODL) (Kohr and Adler, 2017), to generate the X-ray projections. Both the X-ray rasterizer and the 3D voxelizer are differentiable (Zha *et al.*, 2024), allowing the reference-frame CBCT Gaussians to be iteratively updated using gradients from losses defined in the 3D image domain or the 2D projection domain, thereby accommodating various training strategies/stages (see Sec. 2.3.2 for more details).

The spatial component of the motion model is represented using MBC Gaussians, offering a sparse depiction of the MBCs $\boldsymbol{e}_i(\boldsymbol{x})$. Specifically, three spatial levels ($i = 1, 2, 3$) are employed for $\boldsymbol{e}_i(\boldsymbol{x})$ along each Cartesian direction (x, y, z), resulting in a total of nine MBC volumes to be determined. For each spatial level, different numbers of Gaussian points are used for initialization, to capture coarse-to-fine motion. By voxelizing these MBC Gaussians, the time-dependent motion fields, as DVFs, can be derived through the product of the MBCs and their corresponding coefficients, following Eq. 3. Notably, the voxelizer used to voxelize reference-frame CBCT Gaussians and MBC Gaussians are different. Since MBC scores can be negative, we modified the original CUDA Gaussian voxelizer (blue box in Figure 1) (Zha *et al.*, 2024)—which was initially designed to handle only positive values—to a negative-permitting voxelizer (purple box in Figure 1) that supports both positive and negative outputs in the voxelized volume.

To infer the MBC coefficients $\boldsymbol{w}_i(p)$ from a single X-ray projection, we adopt the CNN motion encoder from the DREME (Shao *et al.*, 2025) framework. This encoder is designed to directly extract the coefficients $\boldsymbol{w}_i(p)$ from individual X-ray projections to represent projection(time)-specific motion. Defining motion coefficients based on physical signals like projection-specific X-ray intensity features, rather than a nominal time sequence as in PMF-STINR (Shao *et al.*, 2024), allows real-time motion to be directly inferred from future X-ray scans for motion monitoring. The lightweight CNN encoder comprises six layers of 2D convolutional layers with $3 \times 3$ convolution kernels. The feature maps of these layers consist of 2, 4, 8, 16, 32, and 32 channels, respectively. Each convolutional layer is followed by a batch normalization layer and a rectified linear unit (ReLU) activation function. Following the final ReLU activation, the feature maps are flattened and processed by a linear layer producing nine outputs. Each output channel corresponds to an MBC score $w_{i,k}(p)$, where $i$ represents the three MBC levels and $k$ represents the three Cartesian components, respectively.

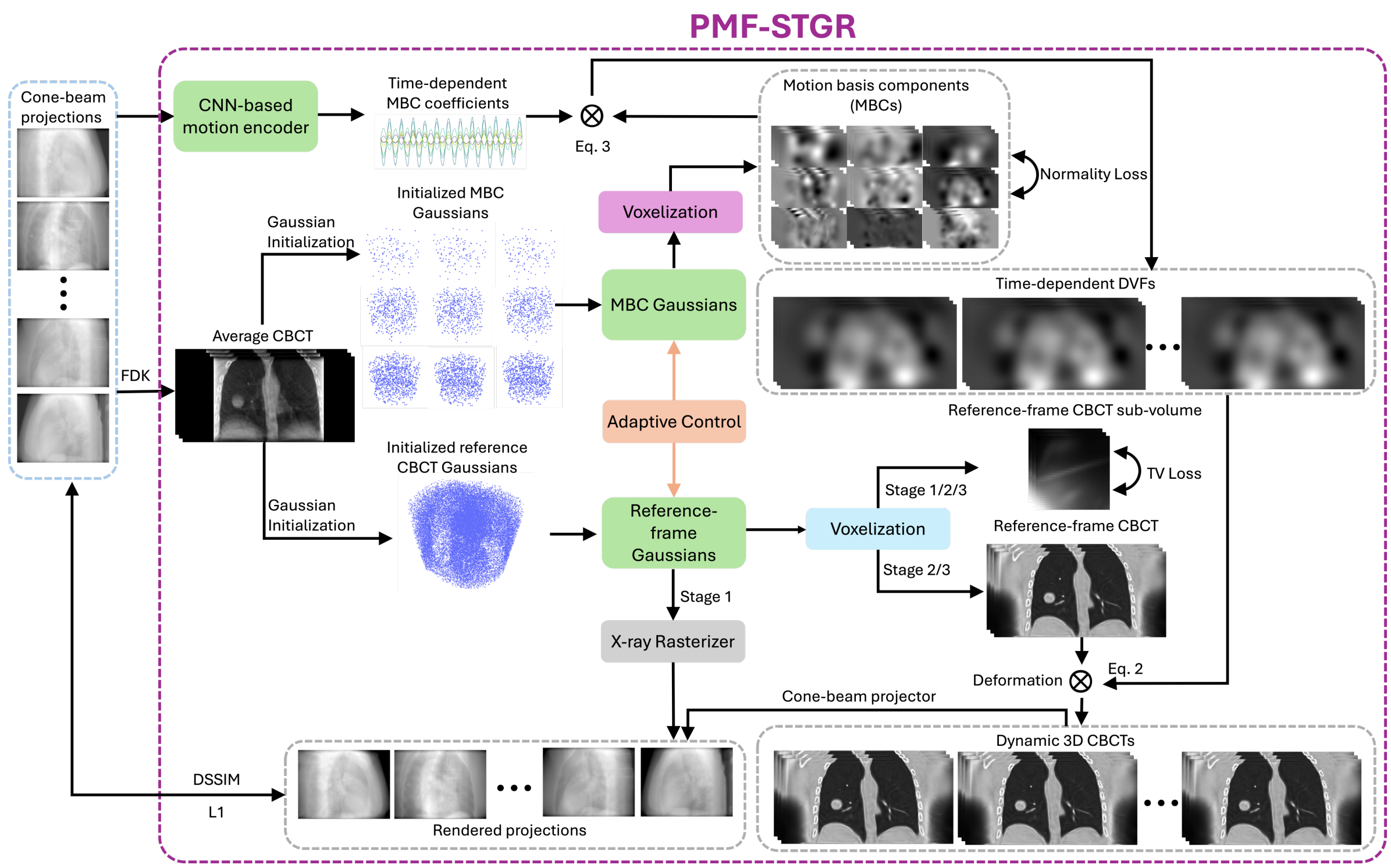

**Figure 1.** Overview of PMF-STGR. Given a sequence of cone-beam projections, an FDK reconstruction is applied to yield a motion-averaged CBCT, which is sampled for the initialization of reference-frame CBCT Gaussians and the MBCs Gaussians. The sampled points of the FDK image are used to position Gaussian kernels. The corresponding FDK image intensities are assigned to the kernels as their initial densities. In training Stage 1, the reference-frame CBCT Gaussians are trained to reconstruct a motion-averaged CBCT with no motion model considered, leveraging a fast X-ray rasterizer to render Gaussians into projections (Gaussian splatting). The training is driven by projection-domain losses (L1 norm loss and a structural similarity loss, DSSIM (Zhou *et al.*, 2004), between the rendered X-ray projections from the CBCT Gaussians and the true X-ray projections) and a CBCT regularization loss (sub-volume total variation loss (Zha *et al.*, 2024)). In training Stages 2 and 3, the reference-frame CBCT Gaussians are voxelized and then deformed into a sequence of dynamic CBCTs using concurrently optimized DVFs, which are generated as the product of MBCs and the corresponding MBC coefficients. The MBCs are obtained by voxelizing the MBC Gaussians with negative-permitting voxelizer (purple), while the MBC coefficients are derived by a CNN-based motion encoder based on the cone-beam projections. Stages 2 and 3 correspond to two levels of dynamic CBCT spatial resolutions (by low- and high-resolution voxelizations of the reference-frame CBCT), to speed up the reconstruction speed and reduce local optima. For both stages, the training of the Gaussians and the motion encoder is driven by projection-domain losses (L1 and DSSIM losses), a CBCT regularization loss (total variation loss), and a motion model regularization loss (normality loss (Shao *et al.*, 2024)). Eventually, a motion-compensated, reference-frame CBCT is solved along with the MBCs and the corresponding projection-specific MBC coefficients to represent the dynamic CBCT sequence.

### *2.3.2 Training strategy*

Based on the to-be-reconstructed cone-beam projections, we employ a quick FDK reconstruction (Feldkamp *et al.*, 1984) to generate a motion-averaged CBCT for Gaussian clouds initialization. To initialize the reference-frame CBCT Gaussians, $M$ points are sampled from the motion-averaged CBCT using grid-based sampling, preserving their corresponding positions and densities. For MBC Gaussian initializations, $M_1$, $M_2$, and $M_3$ ($M_3 > M_2 > M_1$) points are uniformly sampled for three spatial levels to represent coarse-to-fine motion modes. With the negative-permitting voxelizer, we used 9 Gaussians to represent the MBCs, one Gaussian for each spatial level, for three spatial levels and three Cartesian directions.

To improve learning efficiency and mitigate convergence to local optima, PMF-STGR employs a progressive three-stage training strategy after initialization. In Stage 1, we train the reference-frame Gaussians $\boldsymbol{I}_{\mathrm{ref}}$ to reconstruct the motion-averaged CBCT from all projections $\boldsymbol{P}$ with no anatomical motion considered. Digitally reconstructed radiographs (DRRs) are generated via the Gaussian splatting-based X-ray rasterizer (Eq. 9) from the reference-frame Gaussians, and compared with the true X-ray projections with an L1 loss $\mathcal{L}_1$ and a D-SSIM loss $\mathcal{L}_{ssim}$ (Zhou *et al.*, 2004), both defined in the projection domain. The overall loss function $\mathcal{L}_{proj}$ for this stage is given as:

$$\mathcal{L}_{proj}^{1} = \mathcal{L}_1(p_r, p) + \lambda_{ssim}\mathcal{L}_{ssim}(p_r, p), \tag{10}$$

where $p_r$ denotes the rendered projection from the Gaussian splatting-based X-ray rasterizer, and $p \in \boldsymbol{P}$ is the measured projection. $\lambda_{ssim}$ is the weight for D-SSIM loss, which is set to 0.25 based on a trial-and-error empirical search. Since the X-ray rasterizer renders one projection at a time, the batch size is 1 for this training stage with a random projection evaluated for each epoch. In this stage, reconstruction of the reference-frame CBCT is performed at the output (high) resolution level.

In addition to the projection-domain loss, we also incorporated a 3D total variation (TV) regularization loss $\mathcal{L}_{tv}$ to suppress high-frequency image noise while preserving anatomy edges. Following R2-Gaussian (Zha *et al.*, 2024), for each training epoch, we randomly query a sub-volume $\boldsymbol{V}_{tv} \in \mathbb{R}^{D\times D\times D}$ voxelized from the reference-frame CBCT Gaussians for total variation minimization. In this study, $D$ is set to 32. In summary, the loss function for Stage 1 training is defined as:

$$\mathcal{L}_{total}^{1} = \mathcal{L}_{proj}^{1} + \lambda_{tv}\mathcal{L}_{tv}, \tag{11}$$

where $\lambda_{tv}$ is empirically set to 0.05.

In Stages 2 and 3, we train the reference-frame $\boldsymbol{I}_{\mathrm{ref}}$ Gaussians, the MBC Gaussians, and the CNN motion encoder simultaneously, transitioning from coarse (Stage 2) to fine (Stage 3) spatial representations of the dynamic CBCTs for a two-resolution level-based reconstruction. Stage 2 used a coarse resolution (half of that of Stage 3) to voxelize the reference-frame CBCT Gaussians and the correspondingly-deformed dynamic CBCTs, allowing the framework to speed up the reconstruction and better resolve large-scale motion in Stage 2 to reduce the chances of being trapped at local optima. Different from Stage 1, where the DRRs are directly rendered by a Gaussian-splatting-based X-ray rasterizer from reference-frame CBCT Gaussians, Stages 2 and 3 first voxelized the reference-frame CBCT Gaussians into a voxel-based CBCT representation. Based on the voxelized reference-frame CBCT, we applied DVFs derived from the MBC motion model to yield time-resolved dynamic CBCTs, and then used the ODL projector to generate DRRs to compare with the acquired X-ray projections. We did not directly apply a deformation model on the Gaussian clouds, as some other studies did (Wu *et al.*, 2023; Fu *et al.*, 2025), since the deformation model on Gaussians did not represent actual physical motion. Instead, it only represents the motion of the Gaussian kernels, while the true anatomy motion is masked by the compounding effect of Gaussian kernel movement and Gaussian kernel integration/summation. The resulting Gaussian motion fields thus cannot be used towards image-guided radiotherapy applications including contour propagation, tumor localization, or dose accumulation etc. as conventional DVFs. Thus, we chose to solve DVFs based on voxelized CBCTs rather than Gaussians for the better clinical relevance of the former. Similar to Stage 1, for both Stages 2 and 3 the training objective is to maximize the similarity between the DRRs of the voxelized dynamic CBCTs and the corresponding cone-beam projections, under the framework of a motion model. The projection-domain loss $\mathcal{L}_{proj}$ for Stages 2 and 3 is therefore given by:

$$\mathcal{L}_{proj}^{2,3} = \frac{1}{N_{batch}} \sum_{t \in batch} \mathcal{L}_1(\mathcal{P}[\boldsymbol{I}(\boldsymbol{x}, p)], p) + \lambda_{ssim} \mathcal{L}_{ssim}(\mathcal{P}[\boldsymbol{I}(\boldsymbol{x}, p)], p) \,, \tag{12}$$

where $N_{batch}$ is the number of projection samples per batch, and $\mathcal{P}$ denotes the ODL cone-beam projector that generates DRRs from the dynamic CBCT $\boldsymbol{I}(\boldsymbol{x}, p)$. To balance training efficiency and performance, we set $N_{batch} = 32$ for Stage 2 and $N_{batch} = 8$ for Stage 3. In addition, to resolve ambiguities in the spatiotemporal decomposition (Eq. 2) of the low-rank motion model, we incorporate a normality loss to promote MBC normality:

$$L_{MBC} = \frac{1}{9} \sum_{k=x,y,z} \sum_{i=1}^{3} \left( \left( |e_{i,k}|^2 - 1 \right)^2 \right). \tag{13}$$

In addition, Stages 2 and 3 also employ the same TV regularization on voxelized reference-frame CBCT sub-volumes as Stage 1. The loss function for Stages 2 and 3 is then defined as:

$$\mathcal{L}_{total}^{2,3} = \mathcal{L}_{proj}^{2,3} + \lambda_{MBC} L_{MBC} + \lambda_{tv} \mathcal{L}_{tv}, \tag{14}$$

where the weighting factors for MBC regularization $\lambda_{MBC}$ is set to 1. $\lambda_{ssim}$ and $\lambda_{tv}$ use the same values as Stage 1.

To enhance anatomy/motion representations, PMF-STGR incorporates adaptive control mechanisms that dynamically adjust Gaussian distributions during training. Empty Gaussians are removed, while those exhibiting large loss gradients are either cloned or split to increase the representation density. For densification, the density of both the original and newly generated Gaussians is halved, preventing abrupt performance degradation and ensuring the training stability. We implemented PMF-STGR in PyTorch (Paszke *et al.*, 2019), and trained with the Adam optimizer (Kingma and Ba, 2014).

*2.3.3 Evaluation datasets and metrics*

We evaluated PMF-STGR using the Extended Cardiac Torso (XCAT) digital phantom (Segars *et al.*, 2010) and a dataset of lung patient CBCT projections from multiple institutions. The XCAT simulation study served as a source of 'ground truth' for quantitative assessment. In contrast, the patient dataset facilitated an assessment of PMF-STGR's clinical applicability.

**XCAT simulation study.** To evaluate PMF-STGR's performance, we conducted simulations using the XCAT digital phantom, with the imaging field-of-view covering the thoracic and upper abdominal regions. The phantom had a volume of $200 \times 200 \times 100$ voxels, with a voxel size of $2 \times 2 \times 2$ mm³. A spherical lung tumor (30 mm in diameter) was embedded in the lower lobe of the right lung to serve as a motion-tracking target. Seven respiratory motion trajectories (X1-X7) were simulated to assess the accuracy of PMF-STGR in reconstructing dynamic CBCT images under varying motion conditions. X1 represents the simplest case, simulating a quasi-periodic breathing cycle (~5 s) with tumor's center-of-mass moving ~13 mm on average. X2 includes a sudden baseline shift (~5 mm) occurring at the midpoint of the scan (~30 s). X3 introduces variations in both breathing amplitudes and baseline shifts. X4 features a gradually increasing breathing period. X5 simulates a slow breathing scenario, or equivalently a fast-rotation scan in which only a single breathing cycle is captured. X6 combines variations in breathing periods, motion amplitudes, and baseline shifts. X7 has an extended superior-inferior motion range to simulate a large-motion scenario. Based on the dynamic XCAT volumes generated from the motion curves, cone-beam projections $\boldsymbol{p}$ were simulated using the ASTRA toolbox (van Aarle *et al.*, 2016) with a full-fan geometry. The total scan duration was set to 60 s, covering a 360° gantry rotation at a speed of 6°/s. A total of 660 projections were generated at a frame rate of 11 fps to mimic a clinical 3D CBCT acquisition. Each projection was captured at a resolution of $256 \times 192$ pixels with $1.6 \times 1.6$ mm² per pixel.

The quality of the reconstructed dynamic CBCT images was evaluated using the relative error (RE) and structural similarity index measure (SSIM) (Zhou *et al.*, 2004). The relative error was defined as:

$$\mathrm{RE} = \frac{1}{N_p} \sum_{p} \sqrt{\frac{\sum_{i=1}^{N_{\mathrm{voxel}}} \left|\left| \boldsymbol{I}(\boldsymbol{x}, p) - \boldsymbol{I}^{\mathrm{gt}}(\boldsymbol{x}, p) \right|\right|^2}{\sum_{i=1}^{N_{\mathrm{voxel}}} \left|\left| \boldsymbol{I}^{\mathrm{gt}}(\boldsymbol{x}, p) \right|\right|^2}}, \tag{15}$$

where $\boldsymbol{I}^{\mathrm{gt}}(\boldsymbol{x}, p)$ denotes the 'ground-truth' dynamic CBCT corresponding to each projection $p$, and $N_{\mathrm{voxel}}$ denotes the total number of voxels in the image. The accuracy of motion estimation was evaluated by the center-of-mass error (COME) and the Dice similarity coefficient (DSC) of tracked dynamic tumor contours. Specifically, lung tumors were contoured from reference-frame CBCT images and then propagated to the dynamic CBCT instances using the DVFs solved by PMF-STGR. These propagated contours were compared with the 'ground-truth' tumor contours generated via intensity thresholding from the 'ground-truth' dynamic XCAT images to quantify motion estimation accuracy, using the COME and DICE metrics.

**Patient Study.** We further evaluated PMF-STGR using a multi-institutional patient dataset. Table I summarizes the imaging parameters of the study, which included 12 cone-beam projection sets of 8 patients from three sources. From the patient dataset, the MDACC data (P1–P3) were acquired using a Varian system (Varian Medical Systems, Palo Alto, USA) in full-fan mode (Lu *et al.*, 2007). A slow-gantry acquisition covered a 200° scan angle, with scan durations ranging from 4.5 to 5.8 minutes, yielding 1653–2729 projections. The SPARE scan data (P4, P5, P7, P8) were obtained from the SPARE challenge (Shieh *et al.*, 2019), which evaluated 4D-CBCT reconstructions from sparse-view acquisitions in full- and half-fan modes. We selected two full-fan (P4, P5) and two half-fan (P7, P8) patients based on clear anatomical structures that are trackable in 2D projections for motion evaluation. The full-fan and half-fan scans were acquired using an Elekta system (Elekta AB, Stockholm, Sweden) and a Varian system, respectively. For the SPARE dataset, each patient had two sets of projections: a fully-sampled scan and a down-sampled sparse-view scan simulating a 1-minute acquisition (P4-S, P5-S, P7-S, P8-S), where sparse-view scans had much fewer projections shown in Table I. Another in-house UTSW data (P6) was acquired with a Varian system in half-fan mode for about 1 min, covering a 360° scan angle.

Since the patient study lacked 'ground-truth' 3D motion, the accuracy of solved 3D intra-scan motion by PMF-STGR was evaluated in re-projected 2D planes. Specifically, each reconstructed dynamic CBCT was re-projected into a DRR for comparison with its corresponding cone-beam projection. This comparison utilized motion features tracked by the Amsterdam Shroud (AS) method (Zijp *et al.*, 2004), as also used in the PMF-STINR study (Shao *et al.*, 2024). The AS method involves calculating intensity gradients along the superior-inferior direction for both cone-beam projections and DRRs to highlight anatomical landmarks with high-contrast edges for tracking, mostly diaphragms. For P1, as the diaphragm moved out of the field of view, a high-density lung nodule was tracked. For P3, as the diaphragm was indistinct, an alternative high-contrast lung feature was tracked. The gradient image of each 2D projection is then integrated along the horizontal axis for a region exhibiting clear motion-induced intensity variations to form a line profile, and the line profiles of all 2D projections are concatenated to form an AS image. We assessed localization accuracy (LE) to quantify the solved motion accuracy by measuring the differences between the extracted traces from the cone-beam projections and the DRRs. Additionally, we calculated Pearson correlation coefficients between the extracted traces to evaluate their match.

**Table I.** Summary of CBCT imaging parameters of the patient study. The projection size is denoted in width (in pixel number) × height (in pixel number) × Np (number of projections). SAD stands for source-to-axis distance. SDD stands for source-to-detector distance. (-S) indicates the corresponding sparse sampling data of the patient.

| Patient ID | **P1** | **P2** | **P3** | **P4(-S)** | **P5(-S)** | **P6** | **P7(-S)** | **P8(-S)** |
|---|---|---|---|---|---|---|---|---|
| Source | MDACC | | | SPARE | | UTSW | SPARE | |
| Vender | Varian | | | Elekta | | Varian | | |
| Scan mode | Full fan | | | | | Half fan | | |
| Projection size | 512×384×1983 | 512×384×2729 | 512×384×1653 | 512×512×1015 (340*) | 512×512×1005 (340*) | 1024×768×895 | 1006×750×2416(679*) | 1006×750×2918 (677*) |
| Pixel size (mm$^2$) | 0.776×0.776 | | | 0.8×0.8 | | 0.388×0.388 | | |
| kVp/mA/mS | 120/80/25 | | | 125/20/20 | | 125/15/20 | 120/20/20 | |
| SAD(mm)/SDD(mm) | 1000/1500 | | | 1000/1536 | | 1000/1500 | | |
| Reconstructed CBCT voxels | 200×200×100 | | | | | 310×310×102 | 300×300×102 | |
| Voxel size (mm$^3$) | 2×2×2 | | | | | | | |

*The corresponding number of projections of sparse sampling patient cases.

Regarding PMF-STGR's Gaussian initializations, for the XCAT study, we set $M = 50{,}000, M_1 = 20^3, M_2 = 22^3$, and $M_3 = 24^3$. For the patient study, considering the complexity of real patients' anatomy as compared to XCAT, we increased the number of initialization Gaussian points for the reference-frame CBCT ($M = 100{,}000$). For $M_1, M_2$, and $M_3$ we used the same numbers as in the XCAT study. For XCAT and full-fan patient cases, we trained PMF-STGR for 7,000 training epochs, with Stage 1 trained for 5,000 iterations, followed by 1,000 iterations for each of Stages 2 and 3, respectively. For half-fan patients, considering the complexity of half-fan geometry, we added 1,000 iterations for Stage 2 training, resulting in a total of 8,000 training epochs. The other hyperparameters used for training, as described in Sec. 2.3.2, were kept consistent between the XCAT and the patient studies.

We compared PMF-STGR with the state-of-the-art PMF-STINR model. For PMF-STINR, the network architecture and settings were kept the same as originally reported, except that we replaced its MLP-based motion sequencer with a CNN-based motion encoder (Shao *et al.*, 2025) for fair comparison with PMF-STGR.

## 3. Results

### 3.1 The XCAT study results

Figure 2 presents a comparison between reference CBCTs reconstructed using PMF-STGR and PMF-STINR across seven motion scenarios (X1–X7) in both axial and coronal views. Overall, PMF-STGR shows better reconstruction image quality than PMF-STINR with higher SSIM scores (Table II). The reconstructions obtained with PMF-STGR exhibit visibly sharper anatomical structures, particularly around high-contrast regions such as bony structures, as highlighted by the arrows. This improvement can be attributed to the Gaussian-based representation in PMF-STGR, which enables a more structured and adaptive spatial encoding of image features. Unlike PMF-STINR, which relies on an INR and suffers from over-smoothing and blurring in regions with fine anatomical details, PMF-STGR preserves high-frequency features more effectively, leading to improved image fidelity and better structural delineation.

Figure 3 compares the tumor superior-inferior motion trajectories estimated by PMF-STGR and PMF-STINR against the reference 'ground truth' across motion scenarios X1–X7. Both models effectively capture the motion trends; however, PMF-STGR consistently demonstrates improved accuracy with lower COME scores in Table II. PMF-STGR outperforms PMF-STINR in all motion scenarios, achieving more precise trajectory alignment with reduced deviations from the 'ground truth'. For scenario X5, which contains only a single breathing cycle, both models exhibit slight undershooting relative to the reference trajectory. This can be attributed to the limited number of motion states available in training. Nonetheless, the overall tracking accuracy remains high, with PMF-STGR maintaining a lower COME score (0.70 mm) compared to PMF-STINR (0.87 mm), indicating superior robustness in handling sparse motion cycles. The consistent performance of PMF-STGR against varying motion complexities underscores its advantage in 'one-shot' training, making it a more reliable approach for dynamic CBCT motion reconstruction.

Figure 4 compares DVFs estimated by PMF-STINR (top row) and PMF-STGR (bottom row) from the same projection, overlaid on the respective reference-frame CBCTs in coronal and sagittal views. Overall, the DVFs generated by PMF-STGR are more localized and anatomically coherent, capturing finer motion details, particularly in the lung and diaphragm regions. In the sagittal view, the DVFs from PMF-STINR show unrealistic displacements in the spine region, indicating non-physical deformation. In contrast, PMF-STGR, benefiting from its Gaussian-based representation, better captures the sliding motion near lung boundaries and spine, producing more realistic and physically consistent motion.

Figure 5 compares the PMF-STGR resolved dynamic CBCTs with the XCAT phantom 'ground truth'. The case shown is motion scenario X2, containing a 5-mm baseline shift at mid-scan. As shown in Fig. 5, the proposed PMF-STGR reconstruction recovers the motion of both a lung tumor and the diaphragm in the XCAT phantom with high accuracy. The SI motion plots (top of Fig. 5) illustrate that the PMF-STGR trajectory (dashed green) aligns almost perfectly with the 'ground-truth' motion (dashed black). The reconstructed images (bottom of Fig. 5) further demonstrate that PMF-STGR reconstructs dynamic CBCTs with sharp anatomical details. For example, the lung tumor appears as a well-defined high-contrast nodule, and the diaphragm's shape and position are clearly resolved, closely resembling the 'ground truth' at each time instance. Despite some minor mismatches at some air–tissue boundaries when comparing PMF-STGR to the 'ground truth', PMF-STGR resolved the motion correctly and reconstructed dynamic images to match well with the 'ground truth' in general.

For the XCAT study, PMF-STGR achieved an average reconstruction time of ~40 minutes with 17 GB of GPU memory usage for the full-fan scan, compared to ~80 minutes and 30 GB for PMF-STINR, both running on the same RTX 4090 GPU. This demonstrates the efficiency of PMF-STGR over PMF-STINR.

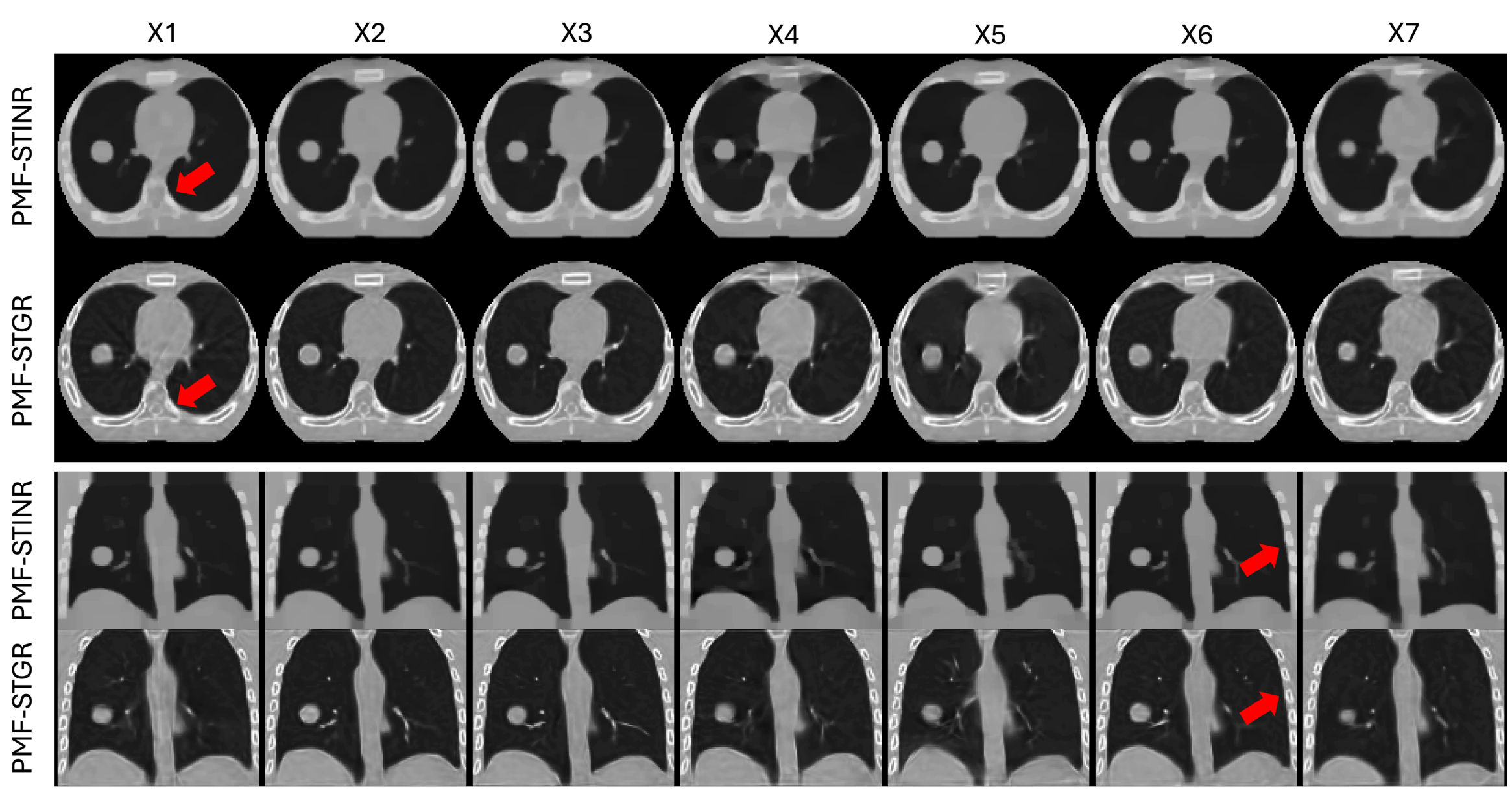

**Figure 2.** Comparison of reconstructed reference-frame CBCTs from seven motion scenarios (X1-7) between PMF-STGR and PMF-STINR. The display window for the CBCT images is 1500 HU at -150 HU level.

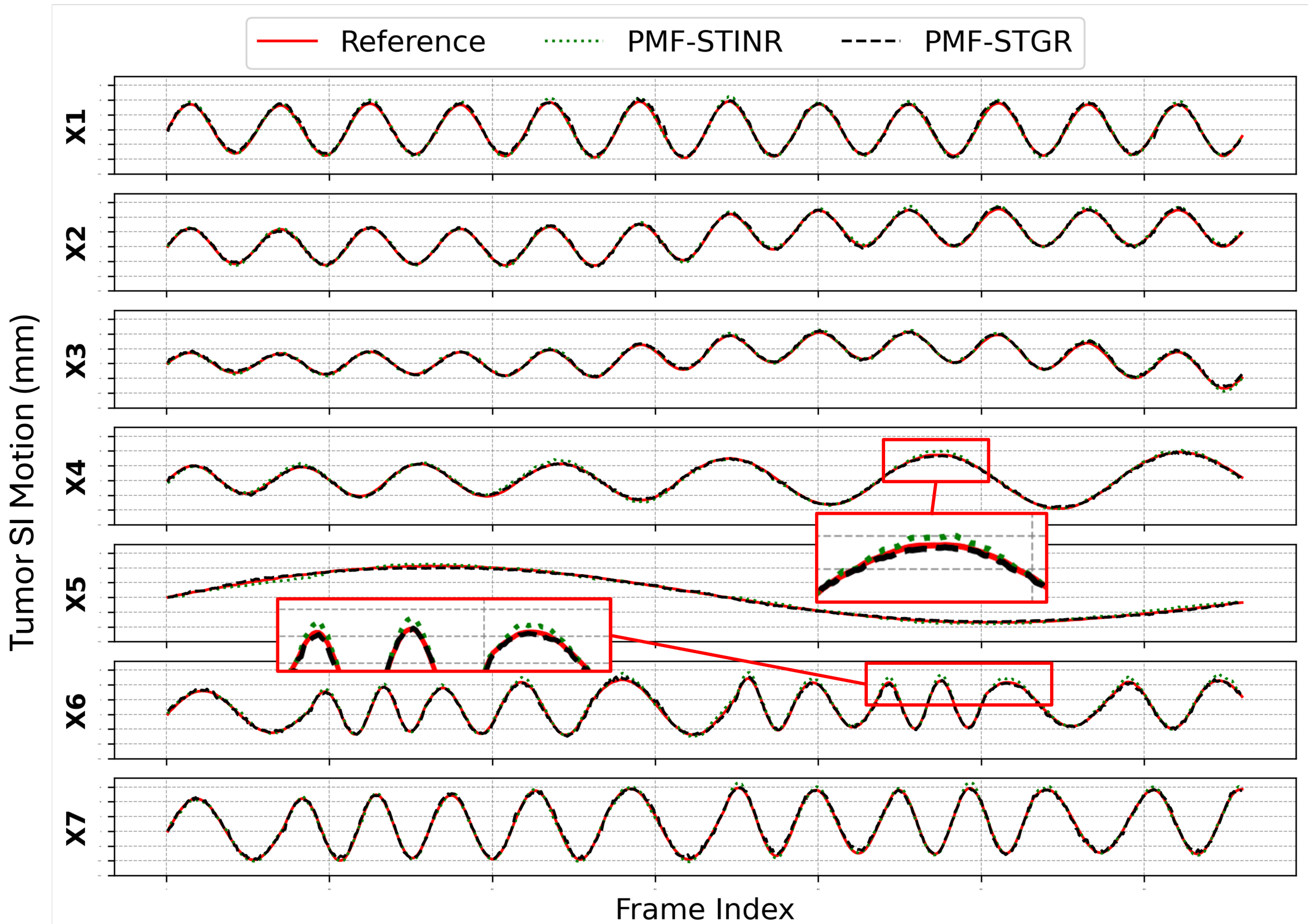

**Figure 3.** Comparison of solved tumor superior-inferior trajectories for motion scenarios X1-7 between PMF-STGR and PMF-STINR, with the 'ground-truth' reference. The red boxes show zoomed-in regions to highlight the trajectory differences.

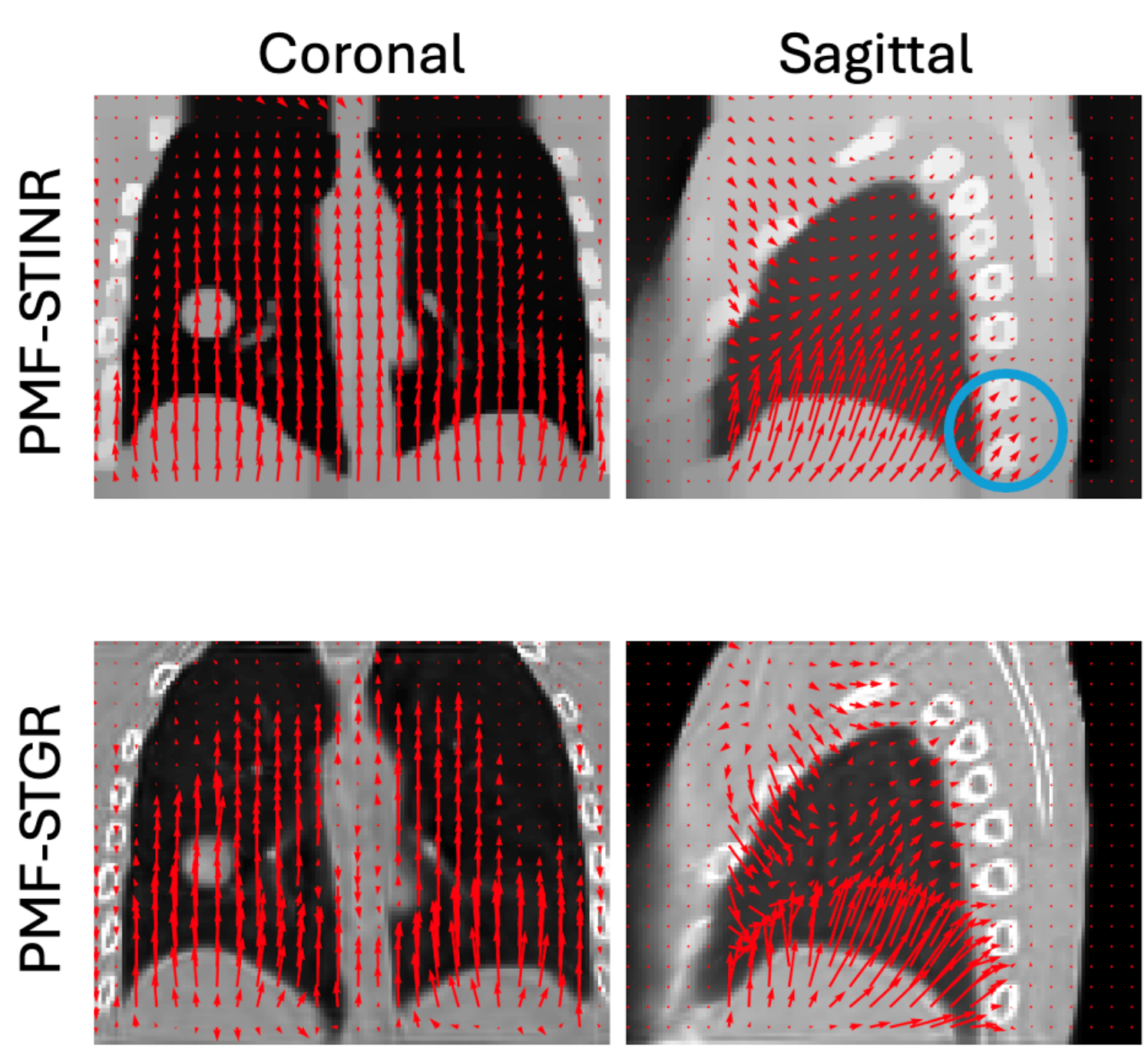

**Figure 4.** Overlays of reference-frame CBCTs and resolved DVFs from PMF-STINR and PMF-STGR on the same X-ray projection (same motion), respectively. Red arrows represent motion fields.

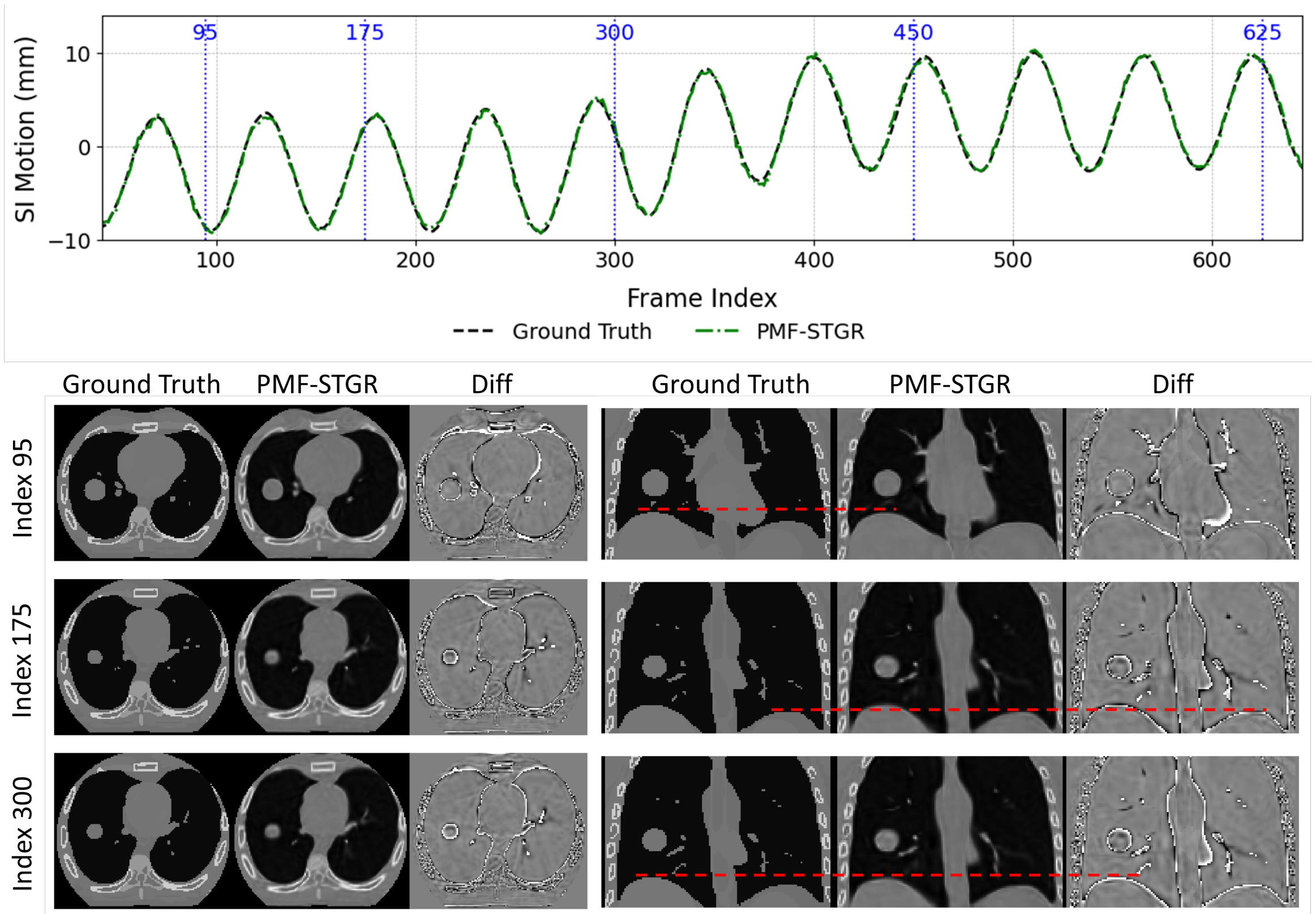

**Figure 5.** XCAT phantom (X2) scenario comparing 'ground truth' and PMF-STGR reconstructions. The first section (row 1) shows the 'ground-truth' and PMF-STGR tumor motion curves along the superior-inferior (SI) direction, with the vertical blue dashed lines indicating the motion states selected for plotting. The second section (rows 2-5) shows the axial (left group) and coronal (right group) views of CBCTs of the selected motion states from XCAT 'ground truth', PMF-STGR reconstructions, and the difference maps between them. The display window for the CBCT images is 1600 HU at 0 HU level, while for the difference images the display window is 600 HU at 0 HU level.

**Table II.** Accuracy of solved dynamic CBCTs and motion for XCAT. The results are presented as the mean and standard deviation (Mean ± SD). Better values are in bold. The arrows are pointing in the direction of higher accuracy.

| Motion | Method | Relative error↓ | SSIM↑ | COME (mm)↓ | DSC↑ |
|---|---|---|---|---|---|
| X1 | PMF-STINR | 0.139±0.011 | 0.949±0.003 | 0.81±0.43 | 0.948±0.017 |
| | PMF-STGR | **0.122±0.006** | **0.991±0.001** | **0.69±0.41** | **0.949±0.015** |
| X2 | PMF-STINR | 0.148±0.003 | 0.933±0.004 | 0.71±0.34 | 0.936±0.015 |
| | PMF-STGR | **0.114±0.006** | **0.992±0.001** | **0.54±0.27** | **0.950±0.010** |
| X3 | PMF-STINR | 0.131±0.002 | 0.944±0.008 | 1.01±0.44 | 0.928±0.019 |
| | PMF-STGR | **0.113±0.016** | **0.992±0.003** | **0.68±0.45** | **0.951±0.016** |
| X4 | PMF-STINR | 0.152±0.005 | 0.942±0.006 | 1.17±1.19 | 0.943±0.022 |
| | PMF-STGR | **0.144±0.012** | **0.987±0.002** | **0.89±0.39** | **0.945±0.014** |
| X5 | PMF-STINR | 0.176±0.004 | 0.941±0.004 | 0.87±0.21 | **0.931±0.047** |
| | PMF-STGR | **0.151±0.009** | **0.986±0.002** | **0.72±0.38** | 0.921±0.020 |
| X6 | PMF-STINR | 0.163±0.004 | 0.949±0.004 | 0.83±0.21 | 0.942±0.017 |
| | PMF-STGR | **0.123±0.005** | **0.991±0.001** | **0.69±0.35** | **0.951±0.012** |
| X7 | PMF-STINR | 0.134±0.005 | 0.949±0.005 | 1.15±0.52 | 0.938±0.024 |
| | PMF-STGR | **0.129±0.009** | **0.990±0.002** | **0.74±0.50** | **0.948±0.018** |

### *3.2 The patient study results*

Figure 6 compares the reconstructed reference-frame CBCTs generated by PMF-STINR and PMF-STGR. Overall, the Gaussian-based PMF-STGR model produces higher image quality, particularly in bone regions, where it reconstructs sharper and more clearly defined structures, as highlighted by the red arrows in Figure 6. This demonstrates the strong representational power of Gaussian models in capturing complex anatomical details. Additionally, the PMF-STGR-reconstructed reference-frame CBCTs of the fully sampled and the sparsely sampled scans (for cases: P4, P5, P7, P8) show comparable image quality, suggesting that the dynamic motions can be captured by PMF-STGR from sparsely-sampled 3D CBCT scans as few as 340 total projections (Table I).

Figure 7 illustrates an example (P1) of dynamic CBCTs reconstructed using the PMF-STGR method. The dynamic motion of the lung nodule is well captured to show lung motion. The first row of the figure displays the superior-inferior motion trajectory, where selected motion states are marked with blue dots. The second through fourth rows present CBCT images corresponding to these selected motion states, where we can observe the motion of the lung nodule. Finally, the fifth row compares the PMF-STGR-derived motion trajectory with the reference trajectory extracted using the AS method, showing strong alignment and validating the reconstruction accuracy.

Figure 8 compares the PMF-STGR-tracked, PMF-STINR-tracked, and reference SI motion trajectories for various anatomical structures, such as the lung nodule and the diaphragm, using the AS image-based method. Both PMF-STGR (black-dashed) and PMF-STINR (blue-dotted) trajectories align closely with the reference motion extracted from cone-beam projections, accurately capturing motion irregularities including amplitude variations, frequency shifts, and baseline drifts. Table III quantitatively evaluates tracking accuracy, showing that both methods achieve sub-millimeter precision, with PMF-STGR exhibiting slightly better performance across all cases. For P5 and P5-S, where the tracked anatomy (diaphragm) is only visible in a subset of the cone-beam projections, both methods successfully infer its motion using other motion features and moving structures within diaphragm-occluded projections. However, since the diaphragm motion cannot be directly extracted as a reference in occluded regions, only the diaphragm-visible section of the trajectory was evaluated.

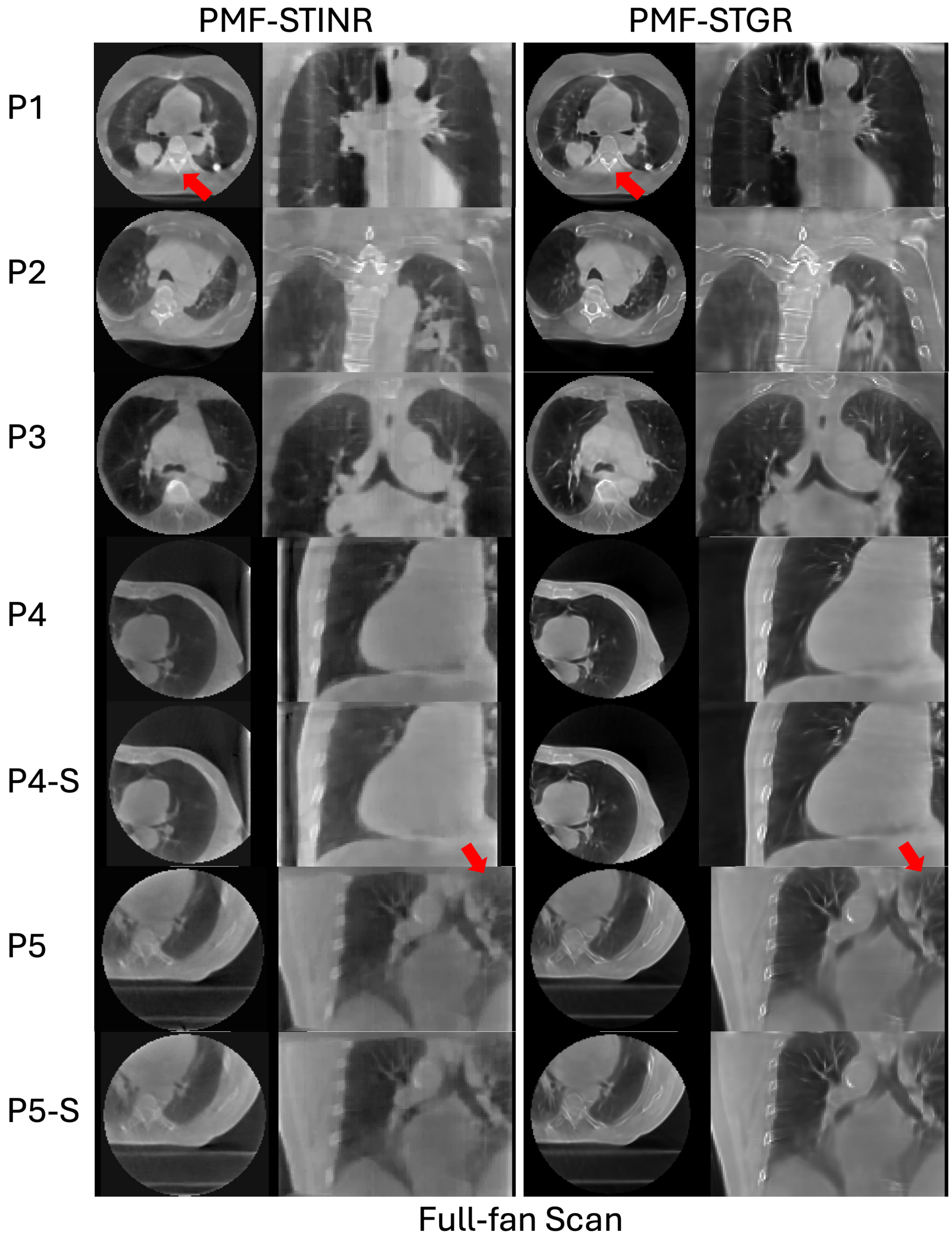

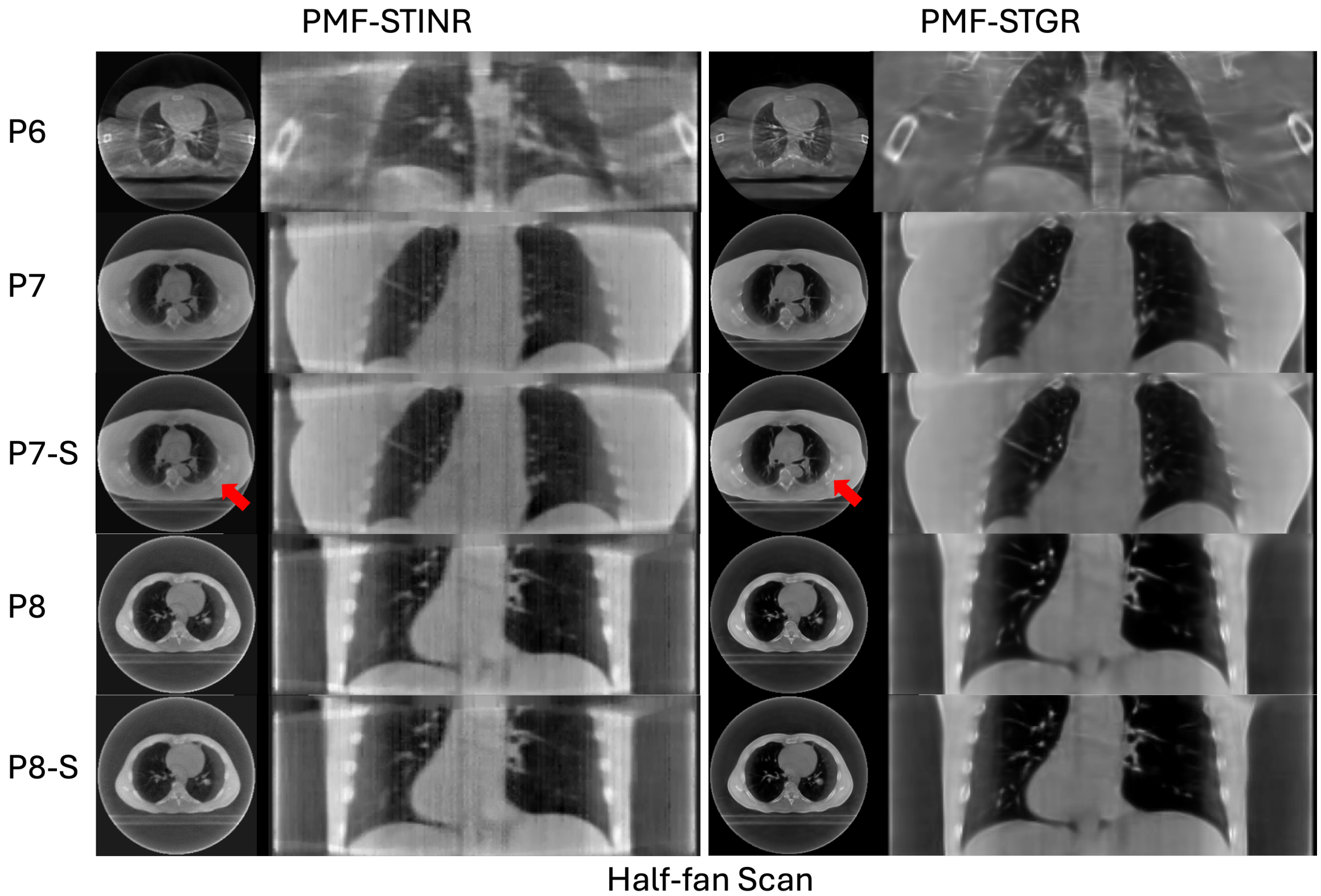

**Figure 6.** Reference-frame CBCTs reconstructed by PMF-STINR and PMF-STGR for the patient study. The display windows and levels for the CBCT images range between 1600 HU and 2300 HU, and between -450 HU and -100 HU, respectively, to optimize the image contrast.

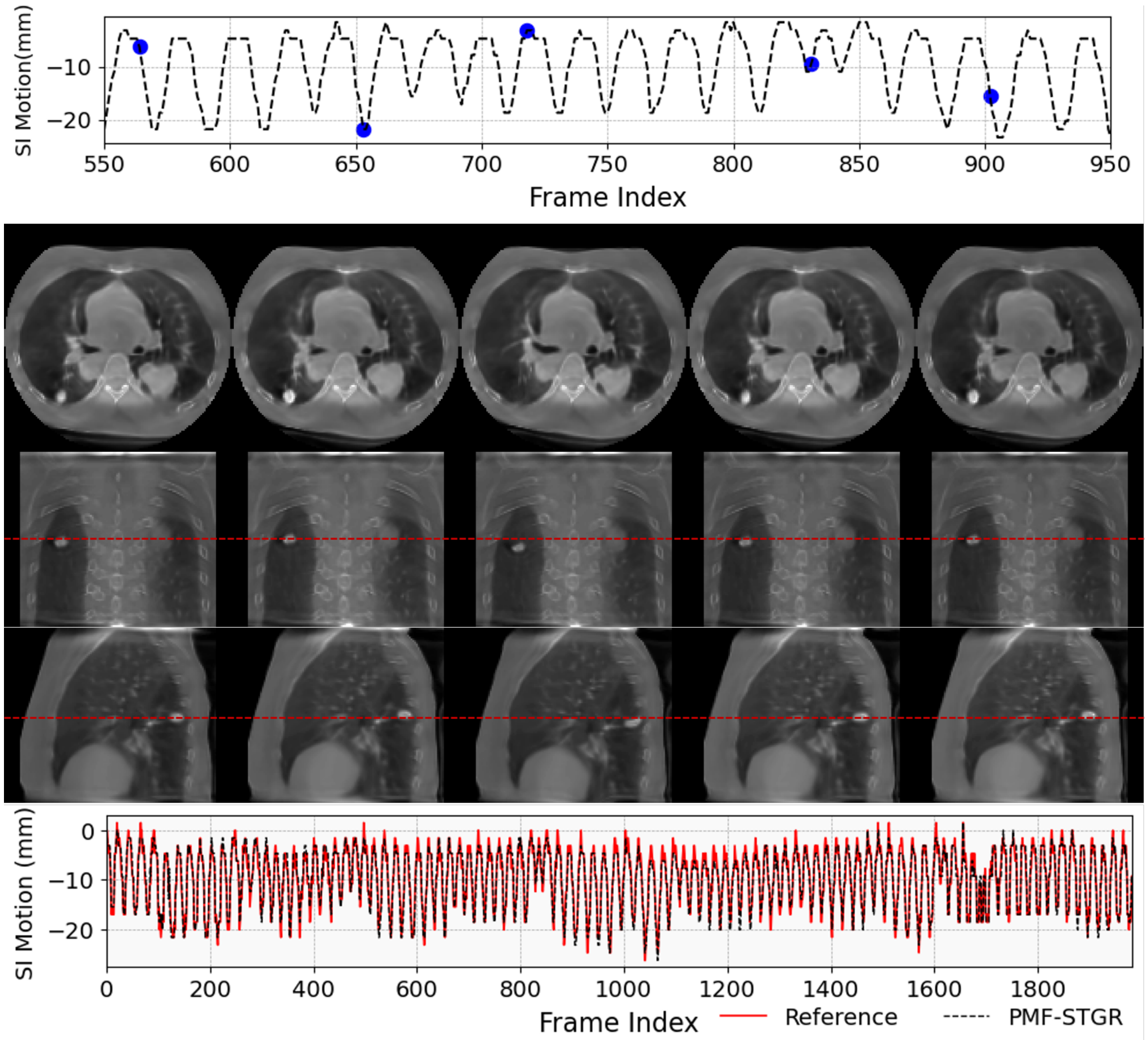

**Figure 7.** PMF-STGR reconstructed dynamic CBCTs for P1. The first section (row 1) shows the corresponding motion curves along the superior-inferior direction, with the blue dots indicating the motion states selected for plotting. The second section (rows 2-4) shows the CBCTs of the selected motion states. The third section (row 5) shows the comparison between PMF-STGR-solved and reference motion trajectories along the superior-inferior direction, extracted using the Amsterdam-Shroud (AS) method. The display window and level for the CBCT images are set at 1400 HU and -120 HU.

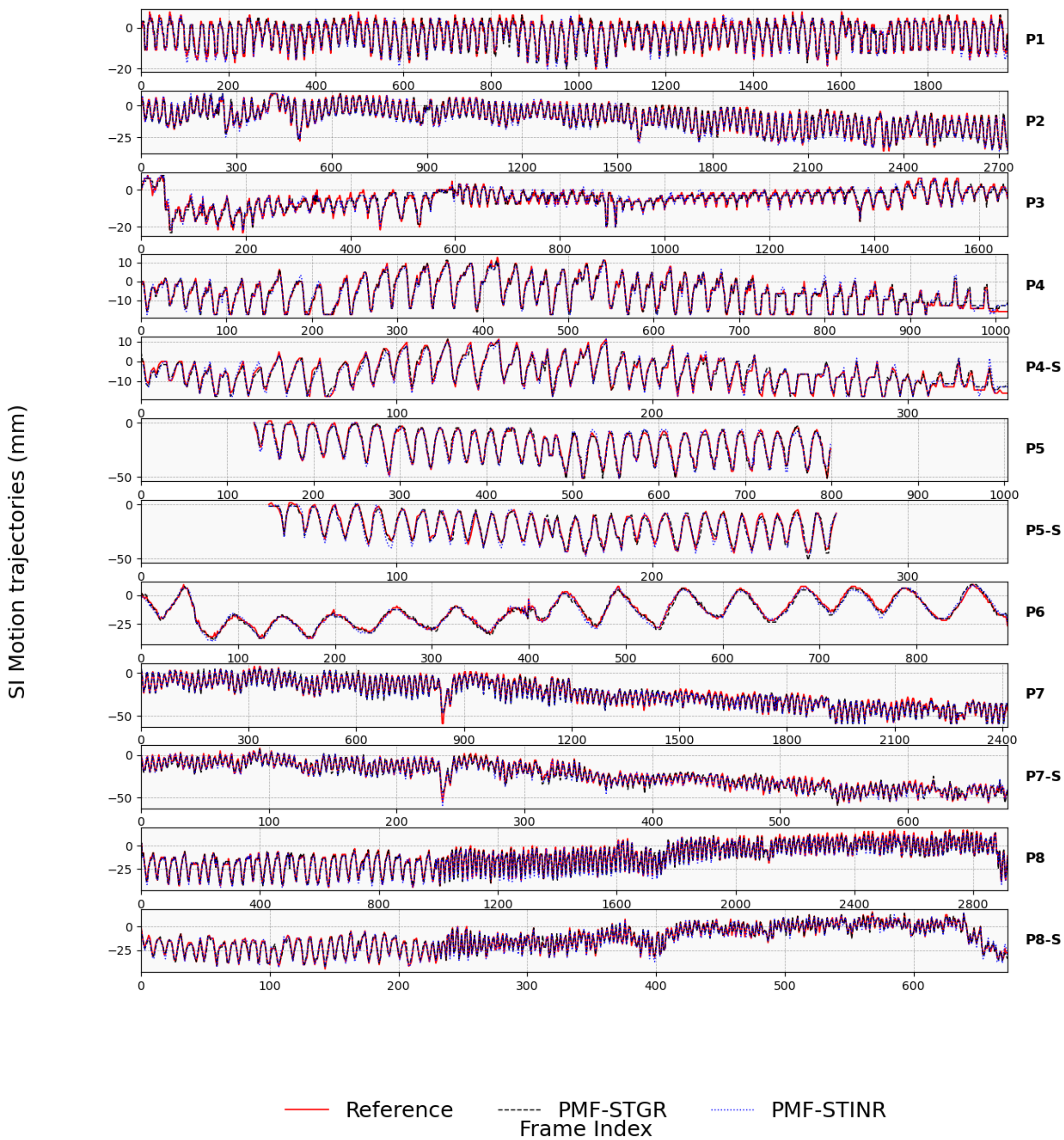

**Figure 8.** Comparison between tracked and reference SI trajectories of P1-P5 for the patient study using the Amsterdam Shroud image-based method, between the PMF-STGR curves, PMF-STINR curves, and the reference curves (extracted from the cone-beam projections). The large change at the end of the P8(-S) tracking curve is caused by the switch of the tracked anatomy for better visibility. The gradual baseline drifts seen in P7(-S) and P8(-S) are caused by changing gantry angles and the associated imaging geometry.

**Table III.** Accuracy of solved dynamic CBCTs and motion for the patient study. The results are presented as the mean and standard deviation (Mean ± SD), where applicable. Better values are in bold. The arrows are pointing in the direction of higher accuracy.

| Patient ID | Method | Pearson correlation coefficient (SI trajectory) ↑ | AS localization error (mm) ↓ |
|---|---|---|---|
| P1 | PMF-STINR | 0.965 | 1.17± 1.10 |
| | PMF-STGR | **0.966** | **1.12±1.08** |
| P2 | PMF-STINR | 0.987 | 1.16± 1.07 |
| | PMF-STGR | **0.989** | **1.10±1.02** |
| P3 | PMF-STINR | 0.955 | 1.07± 1.04 |
| | PMF-STGR | **0.961** | **1.04±1.03** |
| P4 | PMF-STINR | 0.979 | 1.11±1.10 |
| | PMF-STGR | **0.984** | **1.00±1.01** |
| P4-S | PMF-STINR | 0.971 | 1.25±1.15 |
| | PMF-STGR | **0.976** | **1.18±1.09** |
| P5 | PMF-STINR | 0.974 | 2.27±1.93 |
| | PMF-STGR | **0.980** | **2.10±1.43** |
| P5-S | PMF-STINR | **0.977** | 2.27±1.85 |
| | PMF-STGR | 0.974 | **2.17±1.79** |
| P6 | PMF-STINR | 0.987 | 1.48±1.28 |
| | PMF-STGR | **0.989** | **1.33±1.69** |
| P7 | PMF-STINR | **0.989** | **1.68±1.52** |
| | PMF-STGR | **0.989** | 1.70±1.48 |
| P7-S | PMF-STINR | **0.989** | **1.78±1.51** |
| | PMF-STGR | 0.987 | 1.87±1.59 |
| P8 | PMF-STINR | 0.981 | 1.61±1.58 |
| | PMF-STGR | **0.989** | **1.52±1.42** |
| P8-S | PMF-STINR | 0.986 | 1.67±1.49 |
| | PMF-STGR | **0.987** | **1.58±1.46** |

## 4. Discussion

In this study, we introduced PMF-STGR, an innovative framework for dynamic CBCT reconstruction based on Gaussian representations. Unlike previous methods that depend on predefined anatomical or motion models, PMF-STGR simultaneously reconstructs dynamic CBCTs and resolves intra-scan motion directly from cone-beam projections through a 'one-shot' learning approach. This framework addresses the challenging spatiotemporal inverse problem by integrating three main components: a reference-frame CBCT model utilizing a dense assembly of 3D Gaussians, a hierarchical motion model employing coarse-to-fine MBC Gaussians, and a CNN-based motion encoder designed to infer projection-specific motion coefficients. Leveraging the representation power of 3D Gaussians, PMF-STGR enhances both computational efficiency and reconstruction accuracy compared to the INR-based approach, PMF-STINR. Due to the use of CNN-based motion encoder to directly resolve motion from cone-beam projections, the results of PMF-STINR are slightly different from those reported in the original study that used temporal INRs to encode time sequences (Shao *et al.*, 2024). As evidenced in Figures 2-8, and Tables II&III, PMF-STGR achieves high-precision motion tracking while reducing reconstruction time by 50% relative to PMF-STINR, thereby improving its clinical applicability. Moreover, the sparse Gaussian representation decreases memory requirements during model training, in contrast to INR-based methods that map each pixel to represent an entire volume. For reconstructions of full-fan scans at a 2 mm resolution in this study, PMF-STINR's GPU memory usage was approximately 30 GB for a batch size of 32, whereas PMF-STGR required only about 17 GB, approximately halving the memory consumption. For corresponding reconstructions at 1 mm resolution, PMF-STGR consumes about 65 GB for a batch size of 8 (tested on a Nvidia HGX H100 card), while PMF-STINR takes a substantially larger memory to train, with memory use going over the hardware limit (>80 GB) even under a batch size of 1. Additionally, the adaptive capability of MBC Gaussians allows for the better depiction of discontinuous sliding motions of organs against surrounding body walls—a task that poses challenges for B-spline interpolant-based MBCs (Shao *et al.*, 2025), which assume smooth and continuous spatial distributions of MBCs as functions of control points. Compared to pre-defined cubic splines linking control points in B-spline models, Gaussian representations offer greater flexibility in motion description by adaptively splitting, pruning, and cloning during training, making them more suitable for patient-specific, data-driven modeling. PMF-STGR demonstrates robustness across a variety of anatomical structures and complex motion patterns, offering significant advantages for motion-adaptive radiotherapy applications.

Our experiments have demonstrated both quantitatively (Table II) and qualitatively (Fig. 2, Fig. 5, Fig. 6) that the Gaussian-based PMF-STGR produces clearer, sharper anatomical structures than the INR-based PMF-STINR. This observation is attributed to the fundamental differences in how they encode the CBCT volume. Gaussian models use a set of explicit basis functions (Gaussian primitives) to represent the image, which gives them stronger representational power for high-frequency details like bone edges. In contrast, INR-based models, which use a neural network to implicitly encode the image, can struggle to capture fine structural details unless the network is very complex or specially encoded. For example, the original Neural Radiance Field (NeRF) work noted that a plain multi-layer perceptron fails to converge to high-frequency details without techniques like positional encoding to represent high frequencies (Mildenhall *et al.*, 2021). This means INR models tend to blur or smooth out sharp features, whereas Gaussian representations can naturally preserve those features. The improved quality of bony structures in the reference frame CBCT has practical implications for radiotherapy. In the PMF-STGR and PMF-STINR frameworks, the dynamic CBCT frames are generated by deforming the reference frame using solved DVFs. A sharper, more anatomically accurate reference means the computed DVFs better reflect the true motion of tissues/structures more precisely. This directly benefits adaptive radiotherapy, where treatment plans are adjusted based on the patient's anatomy, motion, and accumulated doses. As bony areas are more radiation-absorbing, accurate bone reconstruction and characterization allow the dose to be more accurately calculated and accumulated for dose assessment and adaptive planning.

Although Fig. 3 shows that the SI-motion curves recovered by PMF-STGR and PMF-STINR largely overlap, especially for the smooth trajectories (X1–X3, X7), a closer look at the zoom-in panels of the irregular cases (X5, X6) reveals that PMF-STGR follows the 'ground-truth' trace slightly better. This observation is supported quantitatively by Table II, which reports a lower tumor COME for PMF-STGR than for PMF-STINR. More importantly, the two pipelines were configured with an identical CNN motion encoder so that the only difference lies in the motion representation itself. As a result, PMF-STGR achieves at least the same and sometimes better tracking accuracy while halving GPU memory usage and training time relative to the INR-based PMF-STINR. In other words, the 'marginal' improvement in Fig. 3 should be interpreted as a desirable outcome: PMF-STGR delivers comparable or better reconstruction and motion tracking accuracy at half the computational cost, fulfilling its primary design objective.

While PMF-STGR produces high-resolution spatiotemporal images, we observed in regions that should be uniform (e.g. homogeneous organs), the reconstructed intensity has visible inhomogeneities that are not motion-related (as seen in Figure 5, lung/liver areas). These fluctuations likely stem from non-uniform nature of the Gaussian splatting model. Because Gaussians are instantiated and updated primarily where image gradients are present, broad low-contrast areas tend to receive a sparser, irregular point distribution, which produces voxel-wise intensity variations. Similar behavior, sometimes described as "flattening" of uniform regions, has been observed across several recent Gaussian-splatting-based frameworks (Zha *et al.*, 2024; Fu *et al.*, 2025), indicating that this is a method-level rather than an implementation-specific issue. The clinical significance of this issue depends on the downstream task. Applications that rely directly on CT numbers, such as dose calculation or adaptive replanning, can be sensitive to HU deviations (Schröder *et al.*, 2024), and uncompensated non-uniformity could translate into dose-mapping errors. By contrast, workflows whose primary objective is motion resolution, such as real-time tumor tracking, gating, or DVF-driven deformation, tolerate such intensity variation so long as organ boundaries remain sharp and the deformation field is accurate. Addressing this limitation is an active research focus. One promising direction is the Gaussian-splitting strategy proposed by Feng et al. (Feng *et al.*, 2024b), in which oversized or anisotropic Gaussians are recursively partitioned along an analytically determined plane that preserves the mathematical characteristics of Gaussians. The resulting children Gaussians densify under-sampled regions and equalize local kernel overlap, thereby promoting intensity homogeneity without additional training overhead. Incorporating such uniformity-oriented regularization, or more generally, applying explicit regional-consistency losses, will be an important component of future work to ensure that PMF-STGR reconstructions meet the accuracy requirements of dose-adaptive radiotherapy while retaining their current advantages in motion estimation integrity.

The multi-stage training process of PMF-STGR is designed to handle the challenging spatiotemporal inverse program involved in time-resolved dynamic CBCT reconstruction. In practice, the training is effectively divided into two main parts: (1) Reference CBCT training. We optimize the reference frame Gaussians alone to reconstruct a motion-averaged CBCT from all projections with no motion modeling. (2) Time-resolved CBCT training. We introduce the MBC Gaussians and the CNN motion encoder and train them together with the reference frame CBCT Gaussians in a staged, coarse-to-fine manner to gradually resolve the time-dependent dynamic motion. This sequential training from coarse to fine resolution helps the optimization converge faster and avoids being trapped in local minima when modeling complex deformations. This strategy is analogous to the well-established and widely used multi-resolution image registration approach to improve convergence and accuracy in deformable image registration (Bajcsy and Kovačič, 1989; Vishnevskiy *et al.*, 2017; Zhou *et al.*, 2023; Wang *et al.*, 2024a), where solving a simpler low-resolution problem provides a good initialization for subsequent higher-resolution refinements. Recent studies on Gaussian-based deformable registration (Li *et al.*, 2024a) also validate this approach by employing multi-scale Gaussian primitives to capture both coarse and fine deformations. Additionally, our design of the multiple levels of MBC Gaussians ($M_1, M_2, M_3$) with increasing granularity mimics the classic multi-level B-spline registration strategy to enable a coarse-to-fine refinement of the estimated DVF. This multi-stage, coarse-to-fine scheme enables PMF-STGR to yield high-quality dynamic CBCT volumes. Importantly, even though the training is divided into stages, it is implemented in a streamlined fashion. There is no need to stop, reload, or reinitialize the model between stages: the optimization flows continuously from the reference CBCT training into the motion refinement stages within the same training run, where all components are jointly optimized.

While existing dynamic Gaussian methods (Wu *et al.*, 2023; Lin *et al.*, 2024; Luiten *et al.*, 2024; Fu *et al.*, 2025) focus on deforming Gaussian kernels to incorporate time-varying behavior for motion modeling, our framework instead uses Gaussian representations to model DVFs and perform voxel-based image registration for learning motion deformation. Deforming Gaussian kernels to represent motion is fundamentally limited, as it alters the overlap between Gaussian kernels. Consequently, the image intensity would change inevitably, violating the core assumption of image registration that corresponding anatomical points maintain consistent intensities. This makes Gaussian kernel deformation a non-physical approach to modeling motion. Furthermore, such methods cannot produce DVFs, which are essential in adaptive radiotherapy for auto-segmentation (Rietzel and Chen, 2006; Wang *et al.*, 2008; Xie *et al.*, 2008), organ motion estimation, and dose accumulation. In contrast, our method directly represents DVFs with Gaussians, though it requires a higher GPU memory compared to deforming Gaussian kernel-based techniques. Although PMF-STGR has significantly reduced model training time, a key computational bottleneck remains in the cone-beam projection step within the ODL (Kohr and Adler, 2017). The current implementation is optimized for generating multiple DRRs from a single CBCT in parallel, which suits conventional CBCT reconstruction. However, PMF-STGR requires a distinct DRR for each dynamic CBCT at each gantry angle, necessitating sequential projections across views and introducing inefficiencies. To accelerate reconstruction, future work could implement GPU-based parallelization to compute DRRs from multiple dynamic CBCTs simultaneously, improving throughput and reducing total processing time.

In practice, phase-sorted 4D CBCT images are usually reconstructed within a few minutes to be clinically acceptable (Lee *et al.*, 2016; Mascolo-Fortin *et al.*, 2018). The current PMF-STGR implementation (~40 mins on RTX 4090 for a full-fan scan) is longer than typical clinical 4D-CBCT reconstruction times. However, dynamic CBCT reconstruction is an inherently more challenging task than conventional 4D-CBCT reconstruction. A phase-sorted 4D-CBCT only needs to reconstruct one volume per several hundred projections, whereas dynamic CBCT demands resolving one volume per projection, which dramatically increases the computational burden. This makes the reconstruction of time-resolved CBCT sequences substantially more complex and time-consuming than standard 4D-CBCT, which explains why our current method does not yet meet the clinical benchmark. The primary use of PMF-STGR is image-guided radiotherapy, where dynamic CBCT images are used to monitor patient motion. While it is ideal to reconstruct images as quickly as possible for on-board guidance, not all workflows demand real-time results. For instance, the DVFs solved by PMF-STGR can be used for cumulative dose evaluation and offline adaptive dose planning (Glide-Hurst *et al.*, 2021), based on intra-treatment CBCT imaging concurrent with arc-based delivery. Such adaptive dose re-optimization is typically done between treatment sessions or after a fraction, where a longer processing time is acceptable. In other words, even at its current speed, PMF-STGR could be utilized for off-line adaptive workflows in improving the accuracy of dose delivery or organ-at-risk sparing, which are less time-sensitive.

In routine clinical 4D-CBCT protocols, patient images are reconstructed with anisotropic voxels of roughly $1 \times 1 \times$ 2-3 $mm^3$: the ~1 mm in-plane spacing preserves lateral detail while the thicker (2-3 mm) slice reduces noise when each respiratory phase is built from only a subset of projections (Balik *et al.*, 2013; Lee *et al.*, 2017; Liu *et al.*, 2017). However, in some special cases, for example delineating small tumors in lung SBRT, $1 \times 1 \times 1$ $mm^3$ might be needed (Baley *et al.*). Our experiment shows that a 1 $mm^3$ isotropic voxel grid for a typical patient volume could require ~65 GB of GPU memory, necessitating either GPUs with very large memory (e.g. NVIDIA A100/H100 cards with 80 GB memory), splitting Gaussians across multiple GPUs (Zhao *et al.*, 2024) to train in a distributed manner, or implementing sub-volume-based reconstruction that divides the volume into smaller chunks, reconstructs each chunk sequentially or in a streaming fashion, and then stitches them together (Li *et al.*, 2024b). To further mitigate the memory load, an effective approach is to use a hybrid resolution strategy to decouple anatomical detail from motion detail. For the MBCs, 1 $mm^3$ sampling is not necessarily required to accurately represent breathing-induced motion, since organ motion of a few millimeters can be captured on a coarser grid as DVFs tend to vary smoothly over space. For example, the reference CBCT could be 1 $mm^3$ to preserve detailed structures, while the MBCs are computed on a 2 $mm^3$ grid and up-sampled as needed to deform the reference image, which largely reduces the memory consumption.

Going forward, the training time of PMF-STGR can be further reduced by employing more advanced hardware (faster GPU cards, for instance, RTX 5090). Software techniques like adaptive radius (Wang *et al.*, 2024b) of Gaussian representation can be applied to minimize thread waiting time during the pixel rendering. Additionally, accelerated 3DGS frameworks like FlashGS (Feng *et al.*, 2024a) can be integrated into our Gaussian framework to speed up reconstruction. Beyond pure speed-ups, we can exploit patient-specific priors. Our recently introduced DREME-adapt framework (Zuo *et al.*, 2025) performs a one-time 'virtual-fraction' reconstruction from the pre-treatment 4D-CT, then warm-starts subsequent fractions with the reference CBCT and motion model solved by the preceding fraction in a daisy chain fashion. This adaptation strategy has demonstrated an 85% reduction in training time in our initial tests while maintaining reconstruction and motion tracking accuracy. Practically, using a similar adaptive framework for PMF-STGR could cut the reconstruction time from ~40 minutes down to ~6 minutes, which would potentially meet clinical timing requirements.

Furthermore, PMF-STGR can be further incorporated into a real-time motion estimation framework, such as DREME (Shao *et al.*, 2025). Additionally, the high representational power of Gaussian models can be extended to other modalities such as MRI, as demonstrated by recent research on Gaussian-based MRI representation (Peng *et al.*, 2025), to reconstruct dynamic MRIs.

## 5. Conclusion

In this study, we introduced PMF-STGR, a novel Gaussian representation-based framework for time-resolved dynamic CBCT reconstruction. By leveraging the strong representation power of 3D Gaussians, PMF-STGR enables ‘one-shot’ dynamic CBCT reconstruction from raw cone-beam projections, eliminating the need for prior anatomical or motion models. Compared to the existing PMF-STINR approach, PMF-STGR achieves higher-quality reconstructions with sharper anatomical details, better motion tracking accuracy, and a ~50% reduction in training time, making it more practical for clinical use. Additionally, the sparse Gaussian representation reduces GPU memory requirements while providing a flexible motion model that can better handle discontinuous sliding motions. PMF-STGR represents a promising step toward motion-adaptive radiotherapy, advancing the clinical applicability of dynamic CBCT imaging.

**Acknowledgments**

The study was supported by the US National Institutes of Health (R01 CA240808, R01 CA258987, R01 EB034691, and R01 CA280135) and Varian Medical Systems.

**Conflict of interest statement**

The authors have no relevant conflicts of interest to disclose.

**Ethical statement**

The MDACC dataset used in this study was retrospectively collected from an IRB-approved study at MD Anderson Cancer Center in 2007. The UTSW dataset was retrospectively collected from an approved study at UT Southwestern Medical Center on 31 August, 2023, under an umbrella IRB protocol 082013-008 (Improving radiation treatment quality and safety by retrospective data analysis). This is a retrospective analysis study and not a clinical trial. No clinical trial ID number is available. Individual patient consent was signed for the anonymized use of the imaging and treatment planning data for retrospective analysis. These studies were conducted in accordance with the principles embodied in the Declaration of Helsinki.